\documentclass[aps,prl,reprint,twocolumn,showpacs,floatfix,superscriptaddress]{revtex4-1}

\usepackage{amssymb,amsmath,amstext}                %%   American Physical Society math etc extensions
\usepackage{graphicx}                                               %%   Include figure files                                            
\usepackage{epstopdf}                                               %%   help with eps -> pdf 
\usepackage{color}                                                     %%   change color
\usepackage{bm}                                                        %%   bold math                                    
\usepackage{appendix}                                              %%   formating appendicies
\usepackage[utf8]{inputenc}
\usepackage{bbold}
\usepackage{bbm}
\usepackage{latexsym}
\usepackage{xcolor}
\usepackage{ulem}
\definecolor{lblue} {RGB}{51,71,158}
\usepackage[colorlinks=true,citecolor=blue,linkcolor=blue,urlcolor=lblue]{hyperref}

  \def\be{\begin{equation}}
\def\ee{\end{equation}}
 
\begin{document}
\title{Energy level dynamics across the many-body localization transition}

%%-----  APS (prx) style -----------------
\author{Artur Maksymov}
\author{Piotr Sierant}

\affiliation{Instytut Fizyki imienia Mariana Smoluchowskiego, Uniwersytet 
Jagiello\'nski, ulica \L{}ojasiewicza 11, PL-30-348 Krak\'ow, Poland.}

\author{Jakub Zakrzewski}

\email{jakub.zakrzewski@uj.edu.pl}
\affiliation{Instytut Fizyki imienia Mariana Smoluchowskiego, Uniwersytet 
Jagiello\'nski, ulica \L{}ojasiewicza 11, PL-30-348 Krak\'ow, Poland.}
\affiliation {Mark Kac Complex Systems Research Center, Uniwersytet 
Jagiello\'nski, Krak\'ow, Poland.}

\date{\today}

\begin{abstract}
The level dynamics across the many body localization transition is examined for 
XXZ-spin model  with a random magnetic field.
We compare different scenaria of parameter dependent motion in the system  and consider
measures such as level velocities, 
curvatures as well as their fidelity susceptibilities.  
Studying the ergodic phase of the model we find that the level dynamics does not reveal the
commonly believed  universal behavior after rescaling the curvatures by the level velocity variance. At the same time, 
distributions of level curvatures and fidelity susceptibilities coincide with properly 
rescaled distributions for Gaussian Orthogonal Ensemble of random matrices. Profound differences exists 
depending on way the level dynamics is imposed in the many-body localized phase of the model in which the level dynamics
can be understood with the help of local integrals of motion.
\end{abstract}

\maketitle

\section{Introduction}
\label{sec:intro}
Random matrix theory (RMT) based tools proved very effective in  
statistical analysis of   quantum systems chaotic in the classical limit 
\cite{Bohigas93,Haake}. One of the simplest measures is provided by level 
spacing statistics, Poissonian for integrable systems while revealing 
similarity with RMT predictions for chaotic models \cite{Bohigas84}; notably 
for time-reversal invariant systems solely addressed below, leading to an 
approximate Wigner surmise for the level spacing distribution. 

The same measures proved fruitful for disordered systems. The spectrum of 
{single particle,} Anderson localized system has Poissonian character. In three 
dimensions, in the presence of disorder, systems may become delocalized with 
extended states and Wigner-type level statistics. The 
{Anderson localization transiton}
attracted attention long time ago already, leading to first propositions of 
``intermediate statistics'' \cite{Shklovskii93} being again parallel to similar 
attempts to describe dynamical systems with mixed dynamics \cite{Lenz91}.

With the development of many-body localization (MBL) the statistical 
description of localized and ergodic systems reached a new level. Already in the 
early days of MBL, Oganesyan and Huse \cite{Oganesyan07} introduced a new 
statistical measure, the so called gap ratio, which became a prime tool in a 
statistical analysis of many body spectra. It is defined as 
$r_n=\min\{\delta_n,\delta_{n-1}\} / \max\{\delta_n,\delta_{n-1}\}$ 
where  $\delta_n=E_{n+1}-E_n$ is an energy difference between two consecutive 
levels. 
Being the ratio of two nearest spacings it is a dimensionless quantity, thus it 
may be determined without finding a local mean density of states and 
``unfolding'' the spectra.
As the unfolding procedure is not uniquely defined, 
the introduction of 
gap ratio was a major step {simplifying} statistical description of MBL related 
systems. The mean $\bar r$ appeared as a simple and decisive measure of the statistical 
properties of many-body system with $\bar r\approx 0.53$ corresponding to 
Gaussian Orthogonal Ensemble (GOE) representing ergodic systems while   $\bar 
r\approx 0.39$ corresponding to MBL situation with Poissonian level statistics 
\cite{Atas13}. {Calculation of the average gap ratio $\bar r$ allowed to localize the
transition between ergodic and MBL phases in several models 
\cite{Oganesyan07, Luitz15, Mondaini15,janarek18,wiater18,sierant18,Sierant19b}.}
The analysis of the transition with the more traditional level spacing distribution
and related measures was performed in \cite{Serbyn15, Bertrand16, SierantPRB}.

Level statistics provide, however, only information about $N$ eigenvalues of the Hamiltonian matrix
which contains $N(N+1)/2$ independent matrix elements.
Thus, an additional insight into considered system may be gained by investigation 
of eigenstates.
Their statistical description - e.g. via participation ratios, suffers, however, from 
the choice of the basis used for the analysis \cite{Durt10}.
 Here, as pointed out in a recent analysis of  multifractality across MBL transition
\cite{Mace18} much less is known and open questions exist such as e.g.  a possible existence of non-ergodic yet delocalized 
phase \cite{Pino16,Torres-Herrera17, Luitz17,Pino17} or the question 
of wavefunction properties in MBL regime  \cite{DeLuca13}.
These issues may be addressed by an analysis of generalized 
participation ratios \cite{Mace18}, an {\it alternative}
approach is presented in the present work via 
% 
%For {interacting quantum}
%many-body systems much  less in known about 
the so-called ``level response 
statistics'' which characterizes the sensitivity of individual energy levels 
with respect to   a change in the control parameter.  Edwards and  Thouless 
suggested in 1972 that the conductance of a disordered system may be related to 
the sensitivity of the spectrum to changes of boundary conditions 
\cite{Edwards72}.  This led to level-dynamics studies, the early 
works \cite{Pechukas83,Yukawa85,Nakamura86} established a
link with appropriate Gaussian random matrix theory (RMT) models. In level 
dynamics the role of time is taken by the parameter which is varied. Thus, e.g. 
energy level slopes, $v_n=dE_n/d\lambda$ may be considered as velocities. The 
RMT leads then to  Gaussian distribution of velocities (slopes). The velocity 
distribution ceases to be Gaussian when Anderson localization sets in { -- see}
\cite{Fyodorov94,Fyodorov95v} where distribution of velocities in the localized 
regime has been proposed. 

In the same spirit level curvatures defined as the second derivative 
of energy 
levels with respect to the parameter, $K_n=d^2E_n/d\lambda^2$ were considered 
(we shall not call them accelerations despite {the} motion analogy). 
One of the earliest attempts to describe the curvature statistics for chaotic 
spectra was made by Gaspard in \cite{Gaspard90}. 
The analytical expressions for large curvatures limits  were found  for all 
three Gaussian ensembles: Gaussian unitary (GUE), orthogonal (GOE) and 
symplectic (GSE) ensembles and compared with numerics coming from model quantum 
chaos studies. 
The analytical expressions for curvature distributions were provided for all 
three ensembles on the basis of numerical studies \cite{Zakrzewski93}:
\begin{equation}
P(k)=\frac{N_{\beta}}{(1+k^{2})^{(\beta+2)/2}}
\label{probRMT}
\end{equation}
with $\beta=1,2,4$ for GOE, GUE and GSE respectively, where $N_{\beta}$ is 
normalization constant and $k$ the so called scaled curvature. It has been 
argued \cite{Simons93} that the scaling is universal with $k=K/\gamma$ with 
$\gamma=\pi\rho\sigma_v$ in terms of the mean density of states, $\rho$ and the 
variance of the velocity distribution, $\sigma_v$. The astonishingly simple 
formula (\ref{probRMT}) was proven by von Oppen \cite{vonOppen94,vonOppen95} -- 
for a simple and instructive alternative derivation see also \cite{Fyodorov95}.

Interestingly, the same formula (for GOE) was found to be applicable in the 
case of infinitesimal Aharonov-Bohm flux breaking the time reversal symmetry. 
The numerical findings \cite{Braun94} were confirmed analytically 
\cite{Fyodorov95a}. 

The expression (\ref{probRMT}) describes the typical behavior of curvatures of 
quantally chaotic systems but already there one may observe deviations from 
universality as e.g. for stadium billiard or a paradigmatic model of quantum 
chaos -- a hydrogen atom in a magnetic field. The nonuniversality features were 
related to eigenfunction scarring phenomenon \cite{Zakrzewski93} that modifies 
the small curvatures behavior. Curvature distributions in the Anderson
 localized case and in the transition between extended and localized 
regimes have been addressed in a number of works
\cite{Casati94,Fyodorov95,Canali96,Titov97,Braun97,Zhakereshev99,Evangelou04,
Fyodorov11}. For many-body system with infinitesimal Aharonov-Bohm flux the 
curvature distribution in the MBL regime was studied in \cite{Filippone16}.

Another related measure of sensitivity of a system to a change of a parameter 
is the fidelity, $F$, defined as $F(\lambda)=|\langle \psi(\lambda)|  
\psi(\lambda+\delta\lambda)\rangle|$ (note that often
the square modulus is used for a definition -- see discussion in 
\cite{BZbook}). For small enough $\lambda$ one may expand the fidelity into 
Taylor series $F=1-1/2\chi^2\delta^2+...$ defining the fidelity susceptibility 
$\chi$ (the linear term in the expansion vanishes due to the wavefunction 
normalization). In a standard approach  fidelity and fidelity susceptibility are 
evaluated for a small change of parameter for the ground state wavefunction. The 
latter undergoes dramatic changes at quantum phase transitions (QPT) reflected 
by a maximum of fidelity susceptibility at the critical point (or its divergence 
in the thermodynamic limit  \cite{Zanardi06,You-Li-Gu}). Fidelity susceptibility 
is directly proportional to the Bures distance between density matrices 
corresponding to  $|\psi(\lambda)\rangle$ and   
$|\psi(\lambda+\delta\lambda)\rangle$ \cite{Hubner93,Invernizzi08} -- this 
property can be extended to thermal states \cite{Zanardi07,Sirker10,Rams18}. 
Let us note that universal information can be extracted from fidelity 
susceptibility in the vicinity of the critical points 
\cite{Venuti2007,Zhou08,Schwandt2009,ABQ2010,GuReview,Rams11,Rams11a,Damski13,
Damski14}.

For MBL all states are important so one can introduce fidelity (and fidelity 
susceptibility) of excited states as well (not being limited to thermal 
states). In an interesting approach \cite{Hu16}  it was shown that fidelity of 
specially prepared state (the so called diagonal ensemble) may signal MBL 
transition.
We shall consider an entire fidelity susceptibility distribution for all 
quantum states of the system, also for generic random matrix representations of 
Hamiltonians. Recently, analytic predictions for fidelity susceptibility 
distribution for GOE/GUE dynamics have been derived analytically 
\cite{Sierant19}. We shall consider how this distribution is affected when 
entering then MBL regime. 

For completeness, let us mention yet another measure of level dynamics, the 
distribution of avoided crossing sizes relevant for situations when a change of 
parameter is aimed at being adiabatic. It has been studied in a number of works 
for chaotic systems also in the spirit of RMT \cite{Wilkinson89,Zakrzewski91, 
Zakrzewski93c}. 

As mentioned above, level dynamics provides a complete (via the formalism
of \cite{Pechukas83,Yukawa85,Nakamura86}) access to properties of not only eigenvalues but also
matrix elements in the eigenstate basis of different physical operators. This paves a way to a more
comprehensive understanding of 
MBL which, despite years of studies, remains a controversial
phenomenon \cite{Suntajs19}. The present contribution makes 
the first step in this, hitherto practically unexplored direction.

The paper is organized as follows. First, we provide a 
brief review of level dynamics as typically 
considered in quantum chaos 
studies. Then, we discuss velocities, curvatures, and fidelity susceptibility distributions first 
in the delocalized then in the localized regime. Most surprizingly, we find that in the
delocalized regime the level dynamics indicates signatures of nonuniversal behavior.
In the localized phase we show that the system sensitivity to perturbation is strongly 
dependent on the perturbation itself indicating a connection with description of MBL system
in terms of the so called Local Integrals of Motion (LIOMs) \cite{Huse14,Serbyn13b}.
Finally, we discuss the results and provide future perspectives.
  
\section{The level dynamics revisited}
\label{model}

Let us revisit a simple picture of level dynamics. Consider a general 
Hamiltonian $\hat{H}=\hat{H}(\lambda)$. The particular form often assumed is 
$\hat{H}(\lambda)=\hat{H}_0+\lambda \hat{V}$ but we do not limit ourselves to 
this particular choice. The Schr\"odinger equation  reads
\begin{equation}
\hat{H} |\psi_n\rangle=E_{n} |\psi_n\rangle.
\label{hamld}
\end{equation}
Differentiating this equation side-wise with respect to $\lambda$, and taking 
the left product with $\langle \psi_n|$ we obtain (where the $\dot x$ is a 
substitution for $dx/d\lambda $)
\be
\dot E_n = \langle \psi_n | \dot H | \psi_n\rangle,
\label{velo}
\ee
where $\dot H\equiv dH/d\lambda $ (hereafter the hat is omitted for 
convenience). In a particular case of $H(\lambda)=H_0+\lambda V$ dynamics
suppose that $H_0$ belongs to GOE (GUE). Then for generic $V$ from the same 
ensemble its diagonal elements in the basis of $H_0$ eigenvectors are Gaussian 
distributed giving trivially the Gaussian level ``velocity distribution'' for 
GOE (GUE). 

The situation is more complicated for curvatures $d^2 E_n/d\lambda^2$. 
Differentiating (\ref{velo}) we get 
\begin{eqnarray} 
\nonumber \ddot E_n & =&\langle \psi_n |\ddot H |\psi_n\rangle+\langle 
\dot\psi_n | \dot H | \psi_n\rangle +\langle \psi_n | \dot H | 
\dot\psi_n\rangle \\ \nonumber 
&=&\langle \psi_n |\ddot H |\psi_n\rangle +\sum_k\langle \dot\psi_n | 
\psi_k\rangle\langle\psi_k|V | \psi_n\rangle \\ &+&\langle \psi_n | V |  
\psi_k\rangle\langle\psi_k|\dot\psi_n\rangle,
\label{temp1}
\end{eqnarray}
where we used the resolution of unity. Again differentiating (\ref{hamld}) but 
now  taking the left product with $\langle \psi_k|$ for $k\ne n$ we get
\be \langle\psi_k|\dot\psi_n\rangle=\frac{\langle \psi_k | V | 
\psi_n\rangle}{E_n-E_k},
\ee
which placed in (\ref{temp1}) gives the standard expression for the level 
curvature
\be K_n\equiv \ddot E_n = \langle \psi_n |\ddot H |\psi_n\rangle +2\sum_{k\ne 
n} \frac{|\langle \psi_k | V | \psi_n\rangle|^2}{E_n-E_k}.
\label{curv} \ee
The first term vanishes for $H(\lambda)=H_0+\lambda V$ scenario. In particular, 
for $H_0$ and $V$ belonging to GOE (GUE) ensembles one may easily derive a 
relation between  large curvature tail of curvature distribution, $P(K)$, and 
the spacing distribution, $P(s)$. For generic $V$ the matrix elements in the 
numerator of (\ref{curv}) are independent of the denominator. Large $K$ 
corresponds to small $E_n-E_k=s$. Thus for large $K \propto 1/s$.  If $P(s)$ 
for small $s$ behaves as $s^\beta$ ($\beta=1,2$ for GOE, GUE respectively) then 
$P(K)\propto K^{-(\beta+2)}$ for $K$ large \cite{Gaspard90}.  As mentioned in 
the Introduction after appropriate rescaling $k=K/\gamma$ an exact expression 
for curvature distribution (\ref{probRMT}) is available 
\cite{Zakrzewski93,vonOppen94,Fyodorov95}.

Similarly the fidelity susceptibility may be expressed as
\be \chi_n= \sum_{k\ne n} \frac{|\langle \psi_k | V | 
\psi_n\rangle|^2}{(E_n-E_k)^2}.
\label{fs}
\ee
The analogous argument to that for curvatures \cite{Monthus17} gives the large 
$\chi$ tails of the fidelity susceptibility distribution as 
$\chi^{-(\beta+3)/2}$.

Let us note that the similar arguments gives large curvature and fidelity 
susceptibility tails for integrable (e.g. localized) case where $P(s)$ is 
Poissonian. One may expect $1/K^2$ and $1/\chi^{3/2}$ tails of the 
corresponding distributions assuming that the spacings are independent from 
matrix elements of the perturbation $V$ \cite{Monthus17}.
For curvatures 
it is in apparent contradiction with the log-normal distribution postulated in 
deeply localized regime \cite{Titov97}.

\section{The model}

We consider XXZ model Hamiltonian -- the paradigmatic model of MBL transition 
\cite{Mondaini15} 
\begin{equation}
	H = J\sum_{i=1}^{L-1}\left(S_{i}^{x}S_{i+1}^{x}+S_{i}^{y}S_{i+1}^{y} 
\right)+J_{z} \sum_{i=1}^{L-1}S_{i}^{z}S_{i+1}^{z}+ 
\sum_{i=1}^{L}h_{i}S_{i}^{z} 
	\label{hamXXZ}
\end{equation}
where  $S_i^{\alpha=x,y,z}$ are spin-$1/2$ degrees of freedom at site $i$,  $J$ 
 and $J_{z}$ are the coupling strengths for XY and Z components respectively, 
$h_{i}$  is the  random magnetic field drawn from an appropriate distribution. 
Typically one considers a random uniform distribution in $\left[-W; W \right]$ 
interval where $W$ is the disorder strength. The Hamiltonian (\ref{hamXXZ}) maps 
directly to interacting spinless fermion model:
\be
H_f 
=-\frac{J}{2}\sum_{i=1}^{L-1}\left(f_{i}^{\dagger}f_{i+1}+f_{i+1}^{\dagger}f_{i}
 \right)+J_{z} \sum_{i=1}^{L-1}n_{i}^{z}n_{i+1}+ \sum_{i=1}^{L}h_{i}n_{i}
	\label{hamf}
 \ee
where $f_i$ ($f_i^\dagger$) are fermion annihilation (creation) operators at 
site $i$ with $n_i=f_i^\dagger f_i$ being the occupation at site $i$. In this 
picture $J$ corresponds to the tunneling and $J_z$ is the interaction strength. 
 
\begin{figure}[hb]
	\centering{\includegraphics[width=0.9\columnwidth]{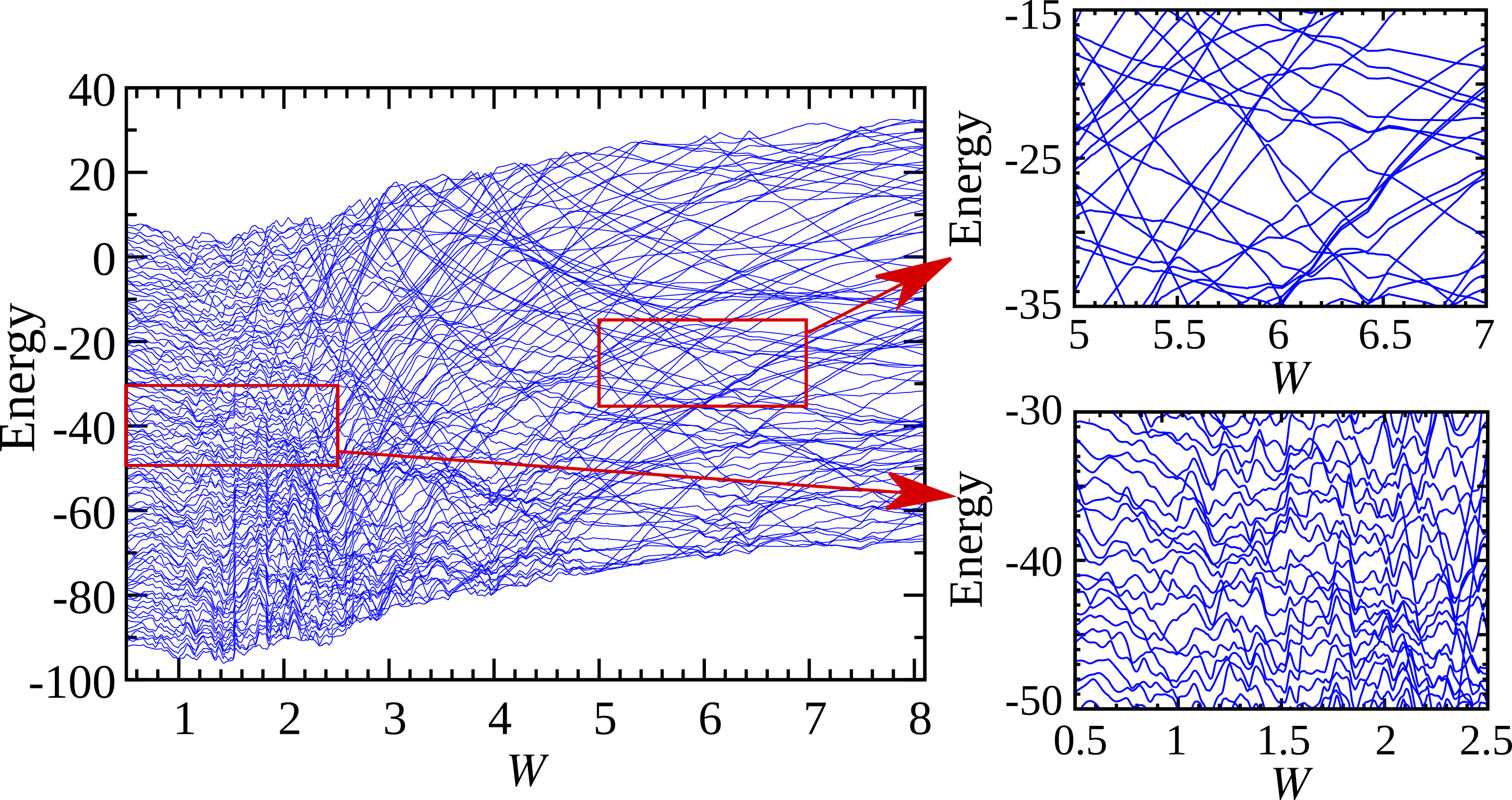}}
	%\vspace{-5ex}
	\caption{Exemplary level dynamics of spin-1/2 Heisenberg model for a 
given single realization of disorder. Both the disorder strength $W$ and the 
energy are expressed in units of $J$. The transition from quantally
chaotic
behavior at low $W$ values  to a regular motion corresponding to many-body 
localization of the system
	at large $W$ is shown. In the former case a number of large avoided 
crossings appears as opposed to large $W$ regime where level crossings are a 
characteristic signature of the presence of constants of motion -- 
 LIOMs \cite{Huse14,Serbyn13b}. }
	\label{fig:diagram}
\end{figure}

There are several ways how one can introduce the parameter change in the 
problem. One may vary the tunneling $J$ affecting XY coupling in the spin 
Hamiltonian. Alternatively, one may modify $J_z$ -- interaction strength in the 
fermion picture. It is also possible to modify random on-site couplings. The example of 
such a change is shown in Fig.~\ref{fig:diagram}. 
Finally, Thouless suggested \cite{Edwards72,Thouless74} that the average conductance is 
proportional to the width of the curvature distribution (where the parameter 
changed, $\phi$, is a small twist  of the boundary conditions).  Explicitly, 
the tunneling term of (\ref{hamf}) takes then the form 
\be\label{twisted}
\sum_{i=1}^{L-1}\left(f_{i}^{\dagger}f_{i+1}\exp(-i\phi)+f_{i+1}^{\dagger}f_{i}
\exp(i\phi) \right).
\ee
Such an approach was applied in the early study of localization in banded 
random matrices \cite{Casati94} and was  frequently used for different systems 
approaching Anderson localization 
\cite{Fyodorov95,Fyodorov95a,Canali96,Titov97,Braun97,Zhakereshev99,Evangelou04,
Fyodorov11}.
For 1D disordered systems the log-normal distribution of curvatures is 
postulated in the localized regime\cite{Titov97,Evangelou04} although the 
deviations from it are indicated in some works \cite{Casati94}. Observe that 
the twist of boundary conditions (passing to the moving frame) explicitly 
breaks time-reversal invariance in the system, so such an approach may exhibit 
different features than changing of $J$ and $J_z$ which keeps the system within 
the same universality class. 

The qualitative picture of the system behavior at different values of the 
disorder amplitude $W$ is shown in Fig. \ref{fig:diagram}. Here and in the 
following all parameters of the problem with the dimension of energy are 
expressed in units of the tunneling $J$ (thus $J=1$, when we study level 
dynamics changing $J$, small changes of $J$ will be considered around this 
\value). The spectrum of a small (for clarity) exemplary system is plotted for a 
single random realization of disorder. Enlarged areas depict two characteristic 
regions. Small $W$ value{s} correspond to delocalized,  ergodic regime with 
a characteristic large number of avoided crossings (with their sizes being 
Gaussian distributed for the appropriate GOE universality class 
\cite{Zakrzewski93c}).  For larger $W$, a transition to a different localized 
regime occurs (where integrability is assured by the existence of Local 
Integrals of Motion (LIOMs) \cite{Huse14,Serbyn13b}). There the avoided 
crossings are replaced by crossings of levels that are an apparent 
manifestation of the existence of LIOMs. 

For our system we shall introduce the following level dynamics

\noindent
{\bf A.} The perturbation  $V$ is given by $H_1=\delta 
J_z\sum_{i=1}^{L-1}n_{i}n_{i+1}=\delta J_z\sum_{i=1}^{L-1}S_{i}^{z}S_{i+1}^{z}$
(the second expression in the spin language) thus interaction strength $J_z$ is 
modified;

\noindent{\bf B.} The perturbation  $V$ is given by $H_2=\frac{\delta 
J}{2}\sum_{i=1}^{L-1}\left(f_{i}^{\dagger}f_{i+1}+f_{i+1}^{\dagger}f_{i} 
\right)$ - thus tunneling term is affected;

\noindent{\bf C.} Twisted boundary conditions  in the form of (\ref{twisted}) 
are assumed.

\section{The delocalized, almost GOE regime}
\label{sec:res}

\subsection{Velocities and Curvatures}
%\subsection{The level dynamics for Gaussian Random Matrix Ensembles}
%\label{subsec:goeLD}
MBL transition occurs in the studied XXZ spin chain 
at $W_C\approx 3.7$ \cite{Luitz15}. As representative value of disorder strength in
the ergodic regime we choose $W=0.5$ for which $\overline r \approx 0.53$, 
staying away from the crossover regime which starts
at $W\approx 2.0$ for system size $L=16$ and also from the integrable point at $W=0$.  
To see how the predictions of universal level dynamics 
\cite{Simons93} are fulfilled let us review its basic findings. 
In essence, the level dynamics should depend on a single parameter -- the 
variance of the level velocities defined as 
$\sigma^2=\langle{(dE/d\lambda)^2}\rangle -\langle(dE/d\lambda)\rangle^2$ where 
$\langle ... \rangle$ denotes average over {disorder} realizations. This implies, in 
particular, that the curvature distributions, regardless of the perturbation 
should be described by (\ref{probRMT}) if rescaled appropriately by the 
velocity variance.

\begin{figure}
	\centering{\includegraphics[width=1.1\columnwidth]{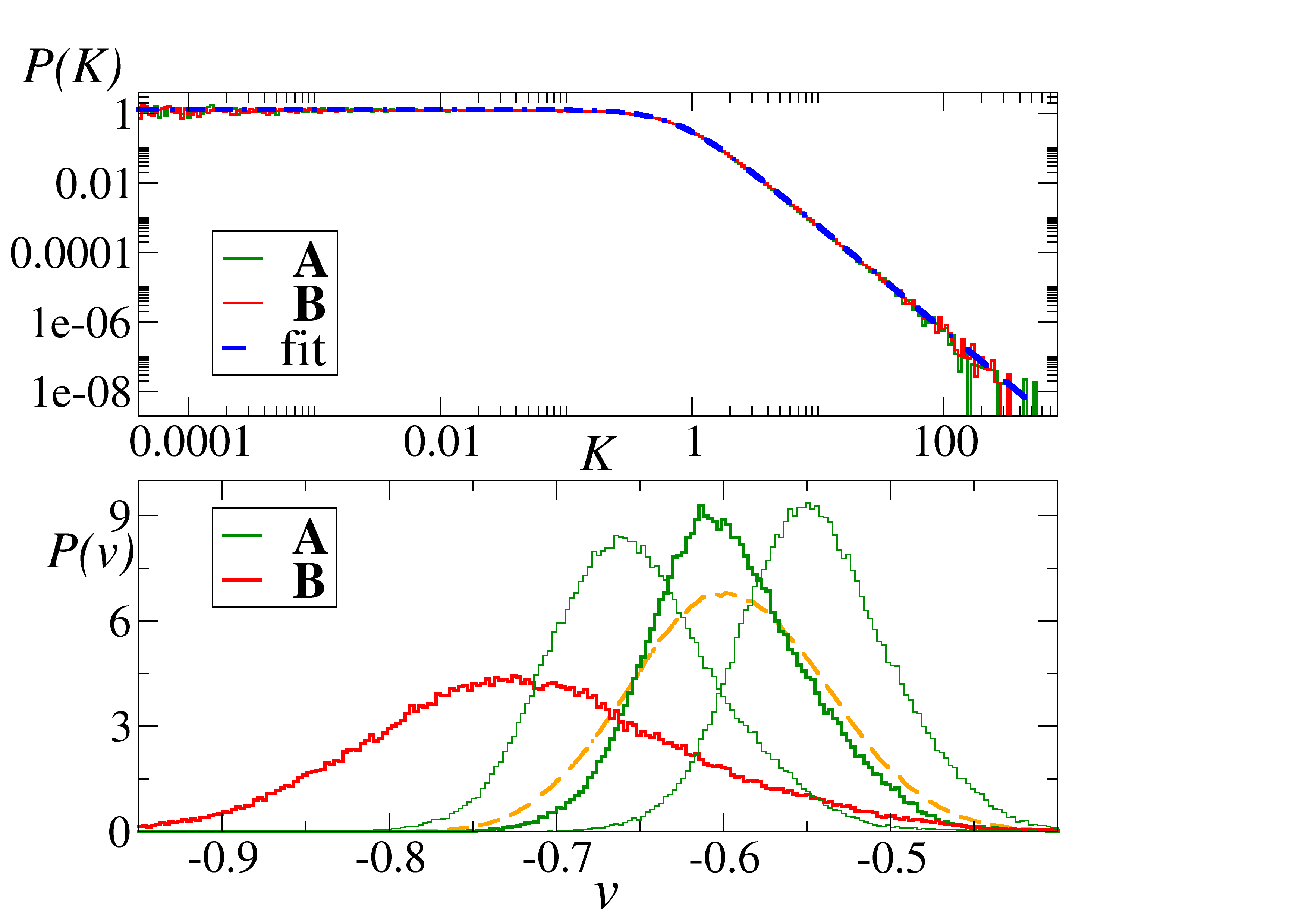}}
	%\vspace{-5ex}
	\caption{Top panel: curvature distributions, $P(K)$, for both types of perturbation assumed 
(see text) coincide.  Both are well fitted with the universal curvature 
distribution (\ref{probRMT}). 
 Bottom panel: the velocity distributions $P(v)$ for the two perturbations are vastly different.
 Shifting the interval from which $n=50$ velocities are taken significantly affects the distribution
 (as shown for perturbation \textbf{A}),
 taking $n=800$ (dashed line) significantly affects the variance of the distribution.} 
	\label{fig:curv}
\end{figure}

Fig.~\ref{fig:curv} reveals that it is not the case. The top panel shows that 
unscaled curvatures $K=d^2E/d\lambda^2$ for both $H_1$ and $H_2$ perturbation{, whereas
the bottom panel shows  the respective velocity distributions $P(v)$.
Firstly, the velocity distributions are not Gaussian as one would expect for GOE but
rather are visibly asymmetric. Moreover, the number of velocities $n$ taken from each 
disorder realization (from the center of energy spectrum) plays unexpectedly important role.
The number $n$ of energy levels for which the average gap ratio remains $\overline r \approx 0.53$, 
is way above $10\%$
of the total Hilbert space that corresponds to $n=1200$ 
eigenvalues from the middle of the spectrum. 
We observe, however, that the velocity distribution $P(v)$ approximates well a distribution for a single
level at the band center only for $n\approx 50$. Shifting the position of the interval from which velocities are 
taken by $2\%$ of the dimension of Hilbert space does not affect $\overline r$ but
results in a significant shift of the $P(v)$ as demonstrated
on example of \textbf{B} perturbation in Fig.~\ref{fig:curv}. Taking $n=800$ velocities from 
the middle of the spectrum results in $P(v)$ 
drawn by the dashed line -- with a significantly larger variance $\sigma^2$. It is thus crucial to
consider $P(v)$ for $n$ not larger than $50$ for $L=16$. This however leads us to an unexpected
result.
}

Since the velocity 
distributions corresponding to perturbations
\textbf{A} and \textbf{B} are vastly different with significantly 
different variances and since distributions $P(K)$ overlap for the two cases,
 the {\it scaled} curvature distributions differ. This 
is a clear indication of a nonuniversal behavior of XXZ Hamiltonian as far as 
level dynamics is concerned in the ergodic regime.
The unscaled curvature 
distribution agreement must be interpreted as accidental. Suppose we reshape the
perturbation $\lambda V$ as $\lambda_1 V_1$ with $V_1=a V$ and 
$\lambda_1=\lambda/a$. Then
unscaled curvatures calculated with respect to $\lambda_1$ are $a^2$ smaller 
then the original ones.
As  the same scaling occurs for the velocity variance the {\it scaled} 
curvatures are not affected by such a transformation. 

Still, in both considered cases, after rescaling properly curvatures $K$ by the 
factor suggested by RMT, i.e. $\gamma=\pi \rho\sigma^2$, the distribution 
obtained does not coincide with (\ref{probRMT}), simply 
$\gamma$ does not yield a proper width. 
Let us note, however, that the distribution is very well reproduced by 
(\ref{probRMT}) provided a width of the distribution is fitted, instead of 
being defined by the velocity variance -- compare Fig.~\ref{fig:curv}.

The surprising,  nonuniversal behavior is robust in the sense that it occurs also 
for systems with different disorder amplitude $W$ (sufficiently small to be far 
from the transition to the localized regime).  Similarly it is robust to 
changes of Hamiltonian. 
{For instance, we have added the next neighbors tunnelings 
assuring that system remains nonintegrable (quantum chaotic) at
$W=0$ (following \cite{Oganesyan07}). This leads 
to a similar, nonuniversal behavior.} On the 
other hand, consistently, we find that the curvature distribution is faithfully 
represented  by (\ref{probRMT}) provided the width of the distribution is 
determined by a fit and not by the velocity variance.

The same distribution works well also for the twisted boundary conditions, case 
{\bf C} as shown already in \cite{Filippone16}. In that case all velocities 
vanish identically due to the symmetry of the Hamiltonian, so the width of the 
distribution remains a sole parameter of the fit. 

\begin{figure}
	\centering{
	\includegraphics[width=1\columnwidth]{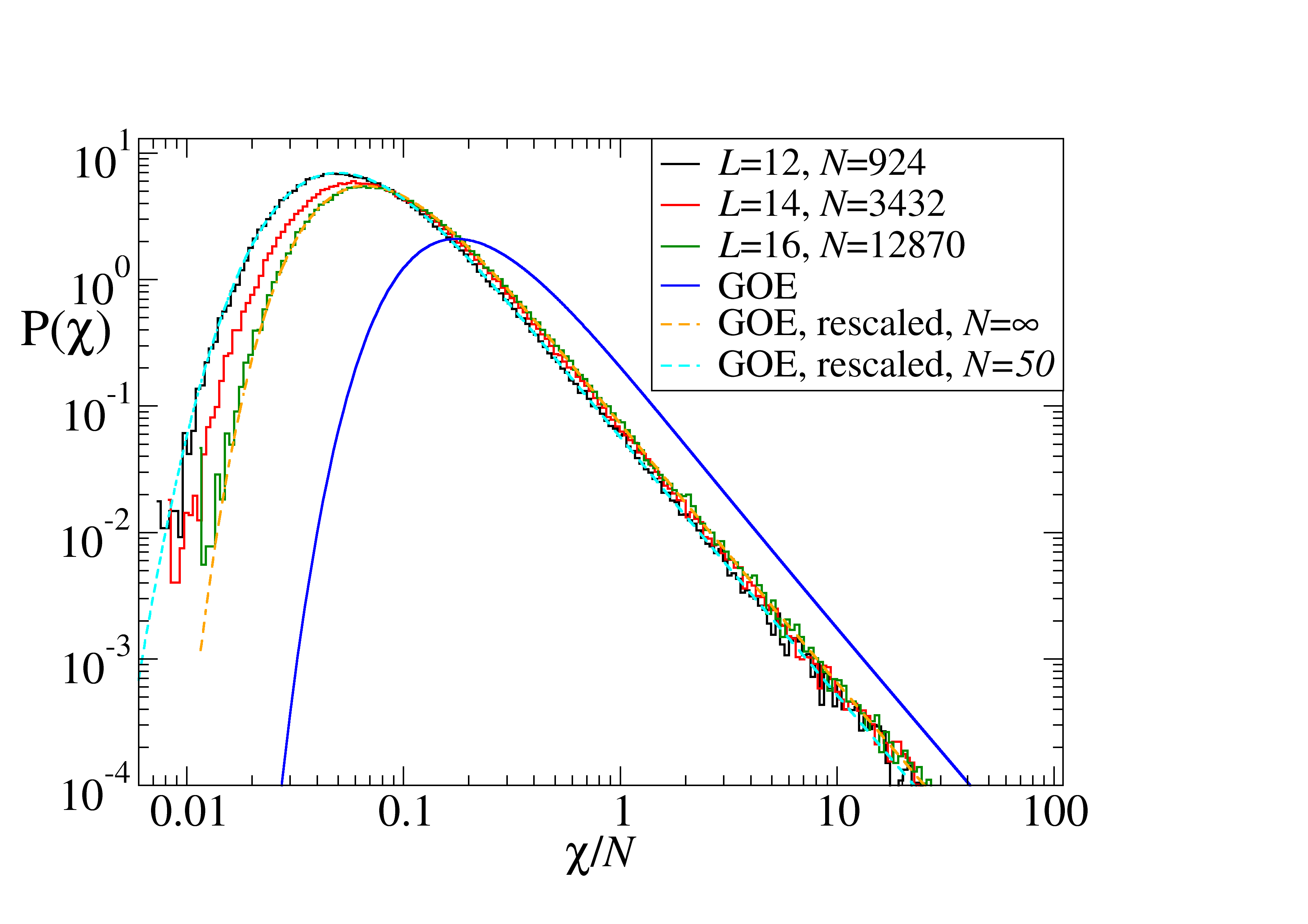}}
	\caption{Fidelity {susceptibility} distribution for perturbation of tunnelings, scheme {\bf B}, 
as a function of the system size in the delocalized regime, $W=0.5$. The large 
susceptibility tails for different system sizes approximately coincides giving 
$x=\chi/N\propto 1/x^2$ in agreement with (\ref{chiunscal}). Small 
susceptibilities show significant size effects {which can be accounted 
for by considering fidelity susceptibility distribution for GOE matrices of finite size
\eqref{eq: 5mt}}. The prediction 
(\ref{chiunscal}) (blue line) satisfactorily reproduces $L=16$ data only after 
rescaling the width, i.e. fitting (\ref{pchi}) distribution.}
	\label{fid}
\end{figure}

\subsection{Fidelity}
\label{fidelgoe}
The fidelity susceptibility distribution has not been studied before for any 
physical system. It seems, therefore, even more interesting to inspect the 
numerical data for this measure for different perturbation schemes. 
Let us recall that there exists an analytic prediction for the fidelity 
susceptibility, $\chi$, distribution, $P(\chi)$ as a function of $x=\chi/N$ for 
GOE and GUE \cite{Sierant19} with $N$ being the matrix rank. 
Fidelity susceptibility distribution for GOE matrices  in 
$N\rightarrow \infty$ limit reads
\be
P(x)=\frac{1}{6x^2}\left(1+\frac{1}{x}\right)\exp(-\frac{1}{2x}).
\label{chiunscal}
\ee
Let us consider the rescaling of the perturbation 
parameter, $\lambda$ in the form $\lambda_1=\lambda/a$ corresponding to the 
perturbation change $\lambda_1 V_1$ with $V_1=a V$.
The very definition of the fidelity susceptibility, via a Taylor expansion 
$F=1-\frac{1}{2}\lambda^2\chi^2$ shows that a transformation to $\lambda_1$ rescales 
fidelity susceptibility by a factor of $a$. The analytic prediction derived 
in \cite{Sierant19} assumes the same density of states both for the original 
Hamiltonian $H_0$ and its perturbation $V$. Since we do not intend to estimate 
the energy scale of the perturbation we  add a scaling factor (playing the role of 
an effective width) defining $y=x\gamma$ with
\be
P_\gamma(y)=\frac{\gamma}{6y^2}\left(1+\frac{\gamma}{y}\right)\exp(-\frac{\gamma
}{2y}).
 \label{pchi}
 \ee
 
As an example consider first the perturbation of tunnelings, i.e. the {\bf B} 
case. The corresponding data are shown in Fig.~\ref{fid} and compared with the 
GOE prediction \cite{Sierant19}. The GOE prediction (\ref{chiunscal}) seems to 
be a bad choice at the first glance but after rescaling  the universal 
prediction, (\ref{pchi}),  represents well the numerical data for the largest 
considered system size, $L=16$. A significant discrepancy at small 
fidelity susceptibilities $\chi$  is visible 
between formula \eqref{pchi} and data for $L=12,14$. To resolve this issue we use the \textit{exact} 
formula for fidelity susceptibility distribution for GOE matrix of size $N$ \cite{Sierant19}
\begin{eqnarray}
 \nonumber 
 P^O_N(\chi) = 
 \frac{C^{O}_N}{\sqrt{\chi}} \left(\frac{\chi}{1+\chi}
 \right)^{\frac{N-2}{2}}\left( \frac{1}{1+2\chi}\right)^{\frac{1}{2}}\\ 
  \left[ \frac{1}{1+2\chi} + 
 \frac{1}{2} \left(\frac{1}{1+\chi}\right)^{2} 
N\frac{N-2}{ N-1/2} \right],
  \label{eq: 5mt}
\end{eqnarray}
where $C^{O}_N$  is a normalization constant, $N$ is assumed to be even.
Defining appropriately rescaled $y=x\gamma$,
we find that the data for $L=12$ are very accurately reproduced if one assumes $N=50$
in \eqref{eq: 5mt}. This result  indicates  that 
\textbf{B} (tunnelings) perturbation of the XXZ spin chain Hamiltonian \eqref{hamXXZ} on $L=12$ 
sites (with Hilbert space dimension $924$) generates the same fidelity susceptibility distribution 
as level dynamics within GOE ensemble of matrices with $N=50$, much smaller than the 
dimension of the Hilbert space. 

\begin{figure}[ht]
\includegraphics[width=1\linewidth]{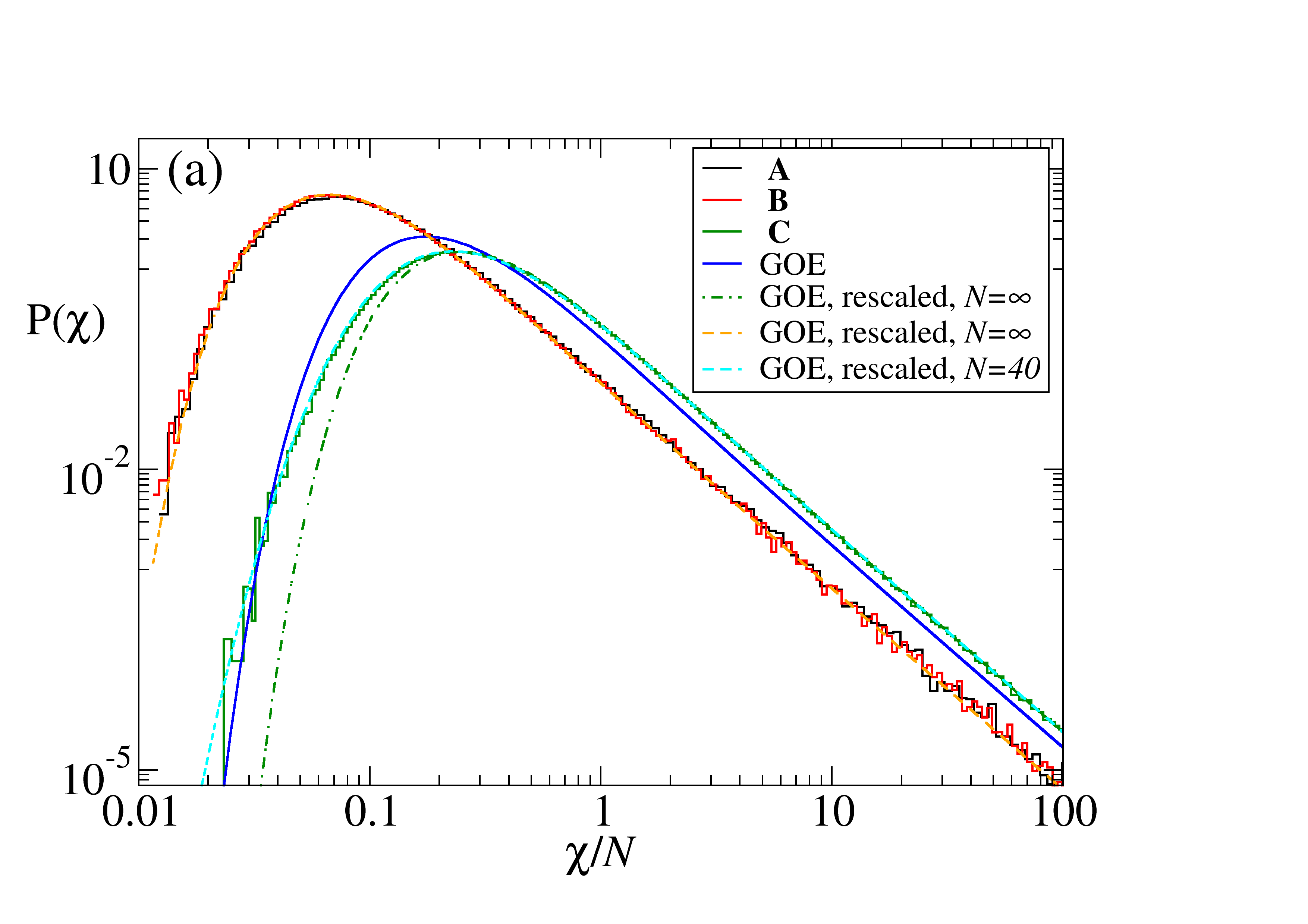} %\\ 
\includegraphics[width=0.9\linewidth]{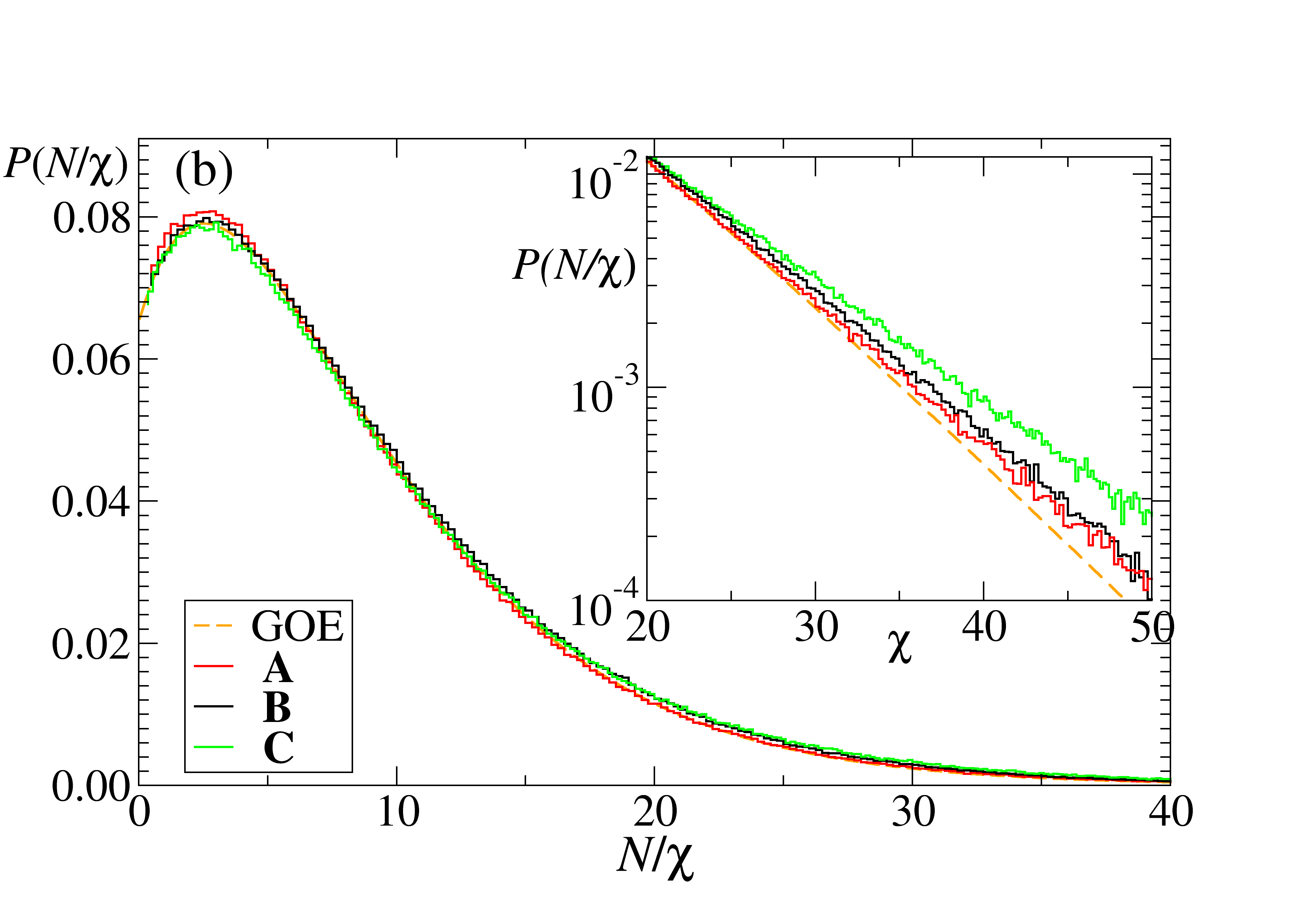} %\\ 
	\caption{(a) Fidelity susceptibility distributions for three types of 
perturbation assumed in the text. While for perturbations of type {\bf A} and 
{\bf B} the distributions coincide, the perturbation corresponding to twisted 
boundary conditions {\bf C} reveals deviations from GOE prediction  for small 
$\chi$. The distributions of the inverse fidelity susceptibility (b) differ mainly
in the tail at large $N/\chi$ and are dependent on the 
type of perturbation assumed.\label{fid2}}
\end{figure}

Fig.~\ref{fid2}a compares the fidelity susceptibility distributions obtained 
for all three perturbations for $L=16$ system. Type {\bf A} and {\bf B} 
perturbations practically coincide, both perturbations preserve time reversal 
invariant symmetry and apparently {the width $\gamma$ reflecting 
the effective energy scale of perturbation} is similar. On the other hand, the green curve in 
Fig.~\ref{fid2} left panel shows the fidelity susceptibility distribution for 
the twisted case {\bf C}. The scaling is remarkably different but this is not 
surprising as the perturbation is vastly different. A
clear deviation 
from (\ref{pchi}) is observed for small fidelities that can be taken into account
by considering \eqref{eq: 5mt} with $N=40$. Note that in this case, the 
ratio of dimension of Hilbert space $(12870)$ and of size of matrices from GOE which reproduce the 
fidelity susceptibility distribution is much larger than for $L=12$ and \textbf{B} perturbation. 
The perturbation {\bf C} 
breaks the time reversal invariance so it is not obvious to what extent (\ref{pchi}) 
and \eqref{eq: 5mt} 
distributions are applicable.

While $P(y)$ has a rather complicated form, (\ref{pchi}) a much simpler 
distribution is obtained looking at the inverse variable, $z=1/y$. Explicitly
\be 
\label{pz}
P(z)=\frac{\gamma}{6}(1+\gamma z)\exp(-\gamma z /2)
\ee
revealing an exponential tail for large $z$ (i.e. small fidelities). The 
suppression of small fidelity susceptibilities in GOE-dynamics is expected due 
to strong level interactions.  Fig.~\ref{fid2}b shows $P(z)$ for all three 
perturbations considered. The corresponding widths are adjusted 
and deviations from GOE-like $P(z)$ behavior \eqref{pz} are observed in the tails
of the distributions.
The differences between resulting fidelity susceptibility 
distributions for the three perturbations are an 
 another manifestation of nonuniversality of level dynamics in the
ergodic regime.

\subsection{Partial summary - delocalized regime}
Before investigating the transition towards MBL it is worthwhile to 
summarize the quite surprizing findings we uncovered in the 
delocalized regime. Contrary to claims \cite{Mace18} that this 
regime is purely ergodic as indicated by the appropriate fractal dimension
obtained from the participation ratio, we have observed a clear breakdown of 
the universality of level dynamics as expected 
for systems faithful to random matrix theory predictions. Matrix elements of
different operators, appropriate for level velocities for different
perturbations, lead apparently to nonuniversal behavior of level
dynamics. Thus, apparently, the considered exemplary system is quite
sensitive to the way in which it is perturbed despite showing eigenvalue
statistics that is in a full accord with GOE.

\section{Transition towards many-body localized regime}

With increasing disorder strength, $W$, the system undergoes a transition
to MBL regime. The transition is expected to occur at $W_C\approx 3.7$ \cite{Luitz15}
in thermodynamically large system. A crossover between the ergodic and MBL behaviors
is observed at finite system size $L$. For relatively small system sizes amenable to 
exact diagonalization, a characteristic value of disorder strength for which 
inter-sample randomness in the system is maximal \cite{SierantPRB} is $W_L\approx 2.7$ for $L=16$ 
(and $W_L \rightarrow W_C$ in the thermodynamic limit).

It is interesting to investigate how different measures of level dynamics 
change across {this} ergodic-MBL crossover. We shall consider here separately 
velocities, curvatures and fidelities -- that will allow, we hope to elucidate 
on the character of the transition as well as on the MBL phase properties.
 
 \subsection{Velocities}
  
Here we may compare only velocities for an ``interaction''  perturbation {\bf 
A} with those for a ``tunneling'' perturbation {\bf B} as velocities for {\bf 
C} case vanish identically due to symmetry considerations. In the deeply 
delocalized regime for small disorder, $W=0.5$ we have seen that both velocity 
distributions were almost Gaussian (with, however, non-negligible skewness, see 
the bottom panel in Fig.~\ref{fig:curv}).

\begin{figure*}[ht]
\includegraphics[width=0.54\linewidth]{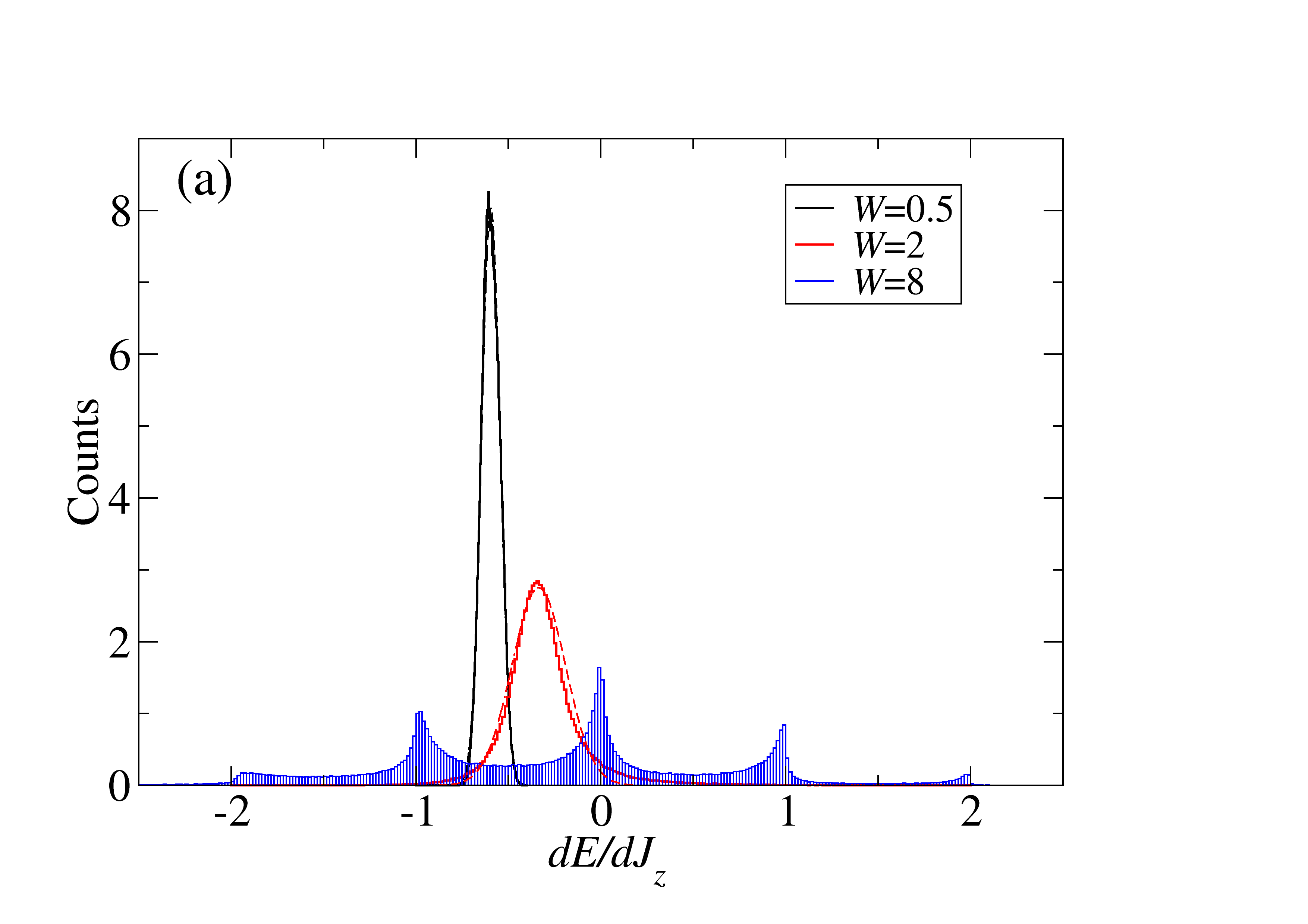}\includegraphics[width=0.53\linewidth]{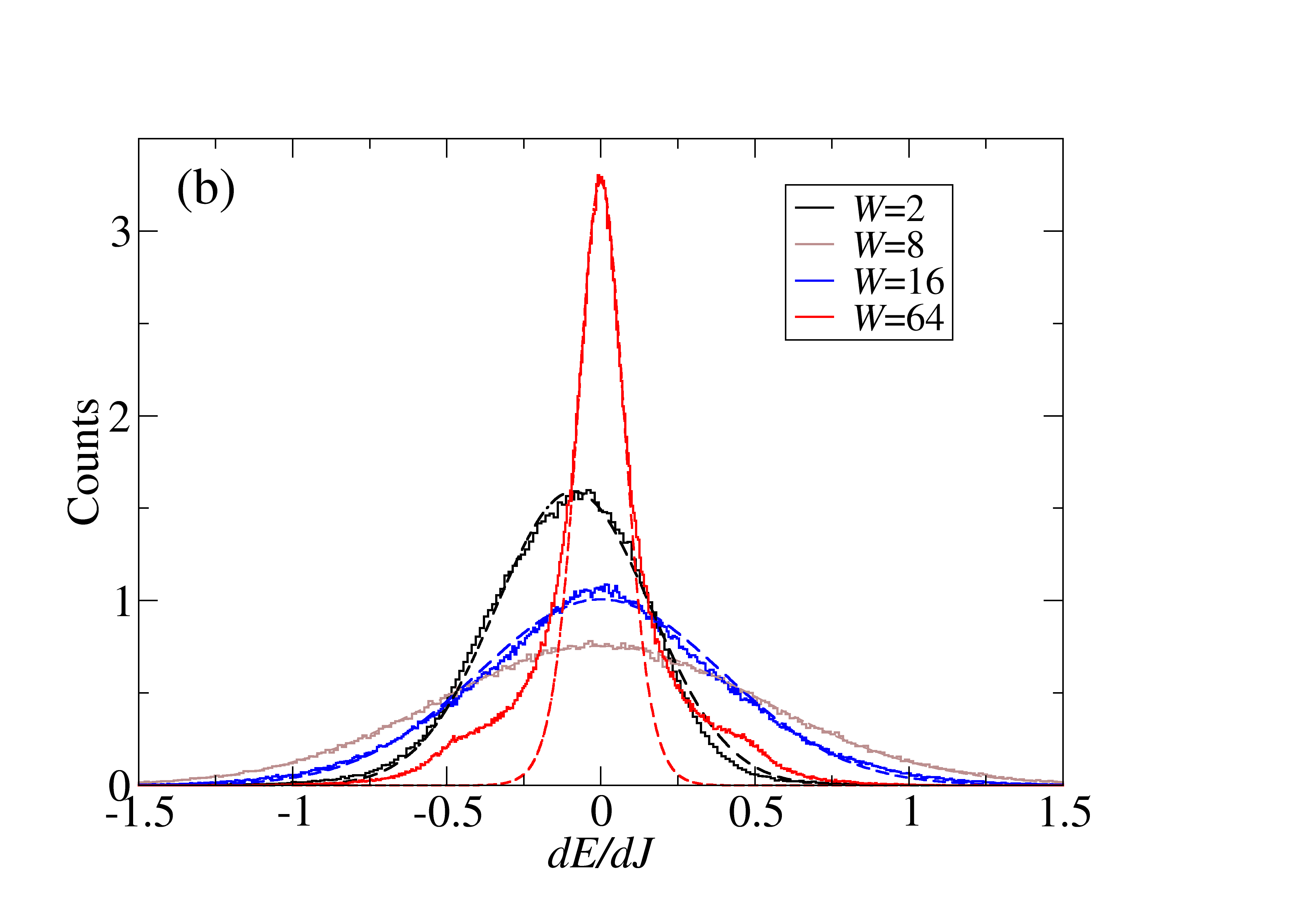} 
	\caption{Velocity distributions for interaction perturbation {\bf A} 
(a) and the tunneling perturbation {\bf B} (b). For intermediate $W$ values (as 
indicated in the figure) relatively narrow distributions for $W=0.5$ are 
broadened and are well approximated by Gaussians. For perturbation {\bf A} in the 
localized regime the distribution develops, surprisingly, a  multi-peak shape -- 
this is a manifestation of LIOMs as discussed in the text. For ``tunneling'' 
perturbation {\bf B} the Gaussian behavior persists even deep in the MBL 
phase ($W=8$). Strong deviations occur in a very deeply localized regime.}
	\label{fig:velo}
\end{figure*}
Fig. \ref{fig:velo} shows numerically obtained distributions of velocities $v$
for both types of 
perturbation. On the delocalized side and across the MBL transition both 
perturbations result in almost Gaussian distributions with widths increasing 
with the disorder strength, $W$. In the MBL phase the velocity distributions 
become markedly different. In particular for ``interaction'' perturbation a  
multi-peak structure appears -- compare Fig.~\ref{fig:velo}a.
While quite suspicious at first glance (with peaks localized approximately at 
integer values of velocities) this surprising structure may be quite easily 
understood taking into account the character of the perturbation which reads, 
recall $H_1=\delta J_z\sum_{i=1}^{L-1}n_{i}n_{i+1}$. Thus velocities are 
nothing else that the average of 
$V=\sum_{i=1}^{L-1}n_{i}n_{i+1}=\sum_{i=1}^{L-1}S_{i}^{z}S_{i+1}^{z}$.
In the MBL phase, there exists a set of LIOMs localized on different 
sites -- those are just ``dressed'' $S_i^z$ operators. Thus $S_i^z$ are 
almost diagonal {in the eigenbasis of $H$}, 
each element of the sum giving approximately $\pm1/4$. Within 
the $\sum S_i^z=0$ subspace for $L=16$ we have eight pairs thus dominantly 
yielding $-2,-1,0,1,2$ as velocities. For slightly smaller system $L=14$ we 
would have then $-3/2,-1/2,1/2,3/2$ as dominant contributions -- and indeed this 
is the result (not shown).

Fig. \ref{fig:velo}b shows the velocity distribution for ``tunneling'' 
perturbation {\bf B}. As in the previous case we observe that in the ergodic 
regime with an increase of $W$ we observe broadening of the velocity 
distribution which is quite well represented by the Gaussian shape. For this 
perturbation the Gaussian shape persists at $W=8$ but going deeper into the 
localized regime we observe first narrowing of the distribution and then strong 
deviations from the Gaussian shape. The distribution for deeply localized 
regime $W=64$ attains quite complicated shape. Interestingly, the central, 
dominating peak may be quite well reproduced (see red dashed line) by the 
velocity distribution  analytically derived \cite{Fyodorov94} in the framework of a nonlinear sigma-model
for a disordered 1D 
wire in the localized regime and given by 
\be 
	P(v)\propto av\frac{(av \coth(av)-1)}{\sinh^2(av)}. 
\ee

\subsection{Curvature distribution}
\label{subsec:cur}

Let us consider now curvature distributions. Immediately, we face the problem of 
an appropriate scaling of curvatures. We shall apply here the idea introduced 
in  \cite{Filippone16}, where
instead of curvature distribution its 
cumulative distribution is concerned. It is defined as
\be 
	F(K)=\int_{-|K|}^{|K|} dK' P(K') \label{integr}, 
\ee
where $P(x)$ is the distribution of unscaled curvatures for a given set. The 
width of the distribution, $\gamma$ is defined as a value of $K$ such that 
$F(K)=1/\sqrt{2}$. It has been found in \cite{Filippone16}
that it is important, in the localized regime, to find the appropriate width 
for {\it each} realization of the disorder, find rescaled curvatures 
$k=K/\gamma$ and then combine distributions from different disorder 
realizations. This reduces the errors \cite{Filippone16} as compared to 
calculating the width of  whole sets corresponding to all realizations. Indeed 
we have confirmed that the difference may be significant in the localized regime 
also for our data, so the results presented are obtained with the former 
approach as in \cite{Filippone16}. For this reason we do not present data for 
Aharonov-Bohm flux case discussed in \cite{Filippone16} commenting only that our 
results are fully consistent with that work. Instead we concentrate on 
perturbations {\bf A} and {\bf B} corresponding to level dynamics in a time reversal invariant system.
 \begin{figure*}[ht]
\includegraphics[width=0.53\linewidth]{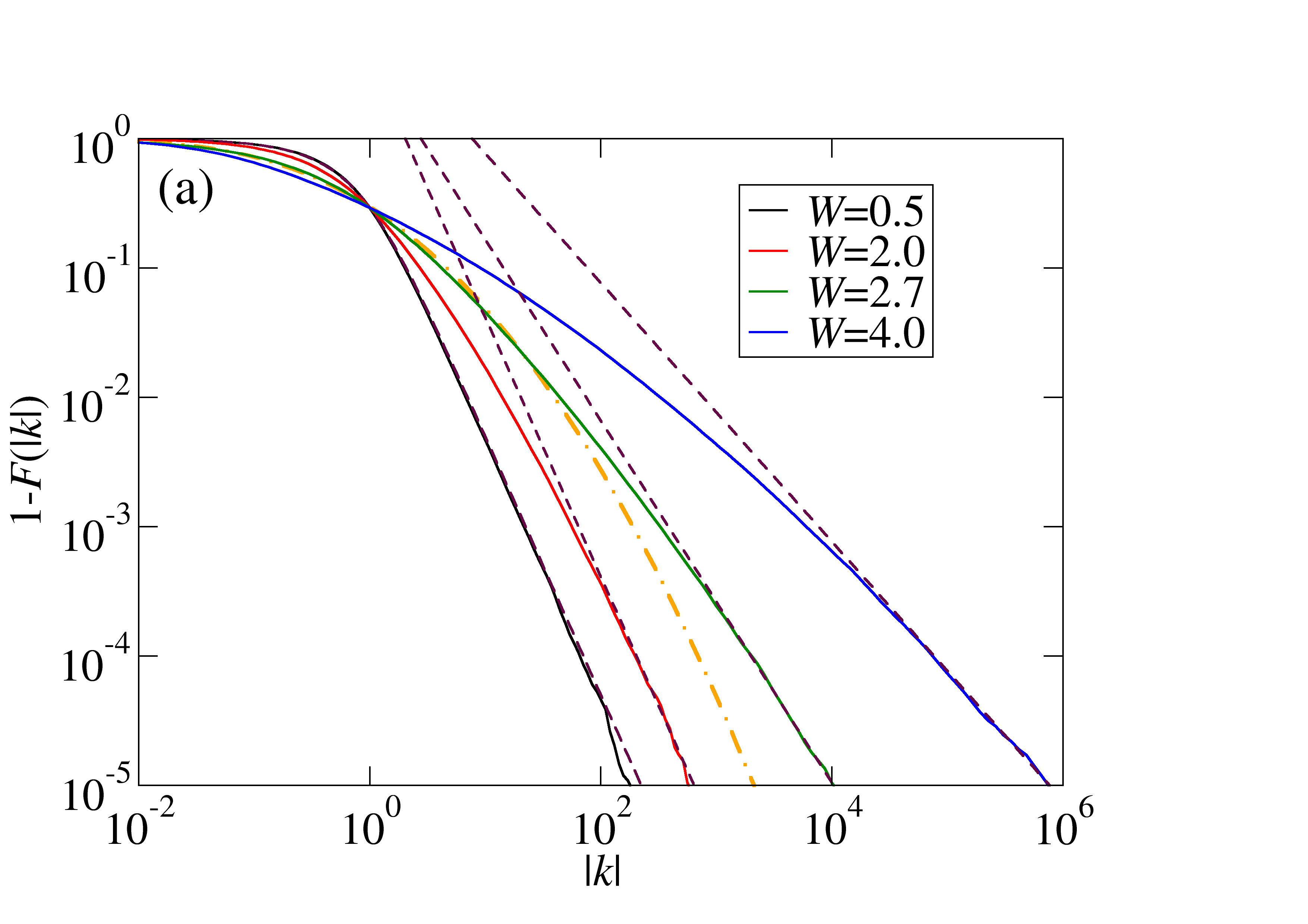}\includegraphics[width=0.53\linewidth]{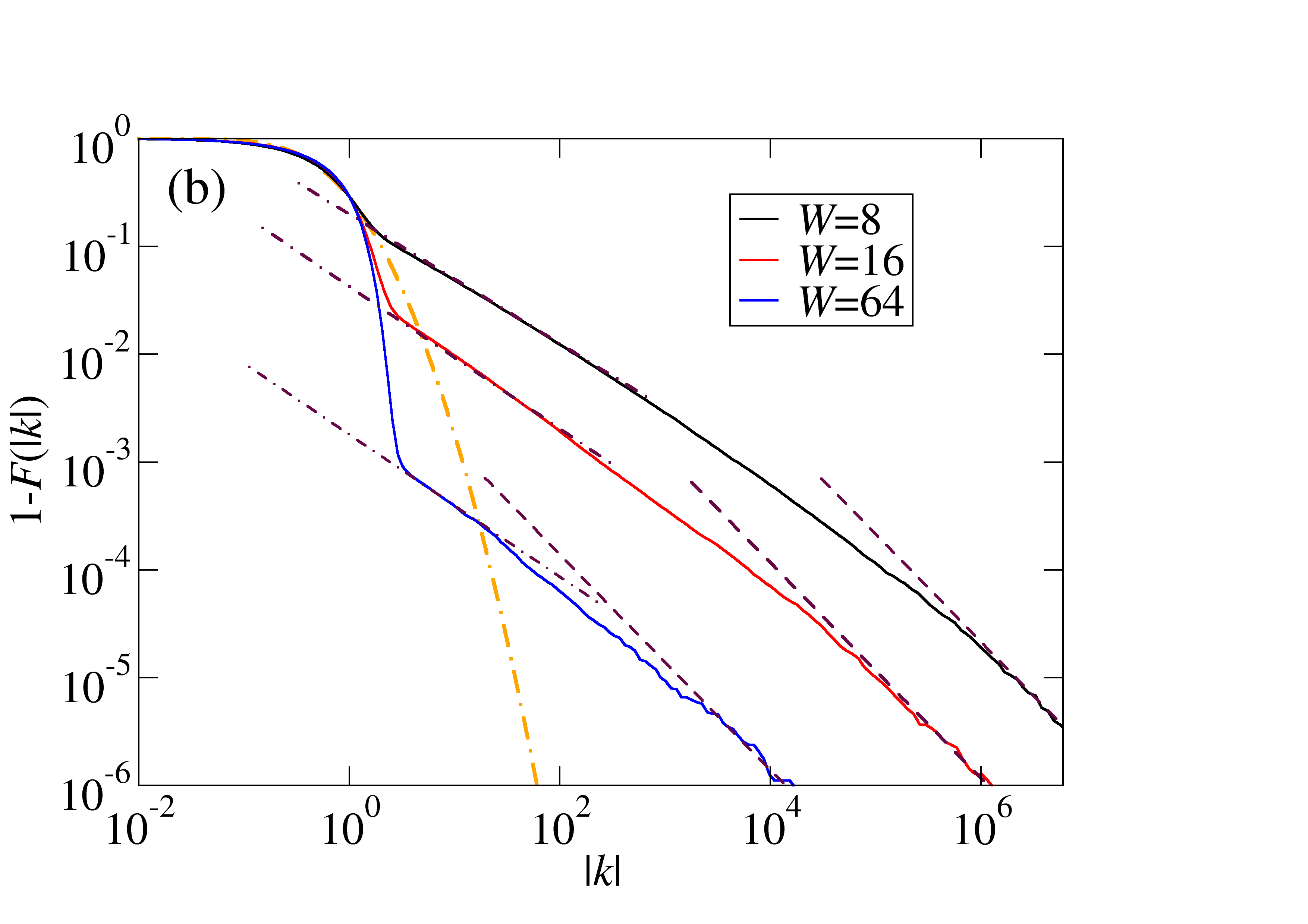} 
 	\caption{(a) The integrated curvature distribution  (\ref{integr}) (or 
rather $1-F(|k|)$) in the transition from extended to localized regime for 
different disorder strengths. Black dashed line corresponds to integrated GOE 
distribution (\ref{probRMT}). With increasing $W$ the  asymptotic power 
behavior {$k^{-\alpha}$} at large $k$ changes smoothly {with 
$\alpha$ ranging from $2$ to $1$} (as shown by fitted dashed lines). The 
orange dash-dotted line gives the log-normal distribution for comparison -- it 
does not fit the data for any $W$ value. (b) The same distributions in deeply 
localized regime (larger $W$). The distribution splits in two component, one 
which exponentially avoids large $|k|$ values revealing approximately Gaussian tails
 $\propto \exp(-gk^2)$
and the second with the algebraic 
tail with $\alpha$ varying smoothly from $0.66\pm0.03$
{to $-1$ at very large $|k|$}. Data 
are obtained for $L=16$ and ``tunneling'' perturbation of {\bf B} type.}
 	%	\vspace{-3ex}
 	\label{fig:curvat}
 \end{figure*}

Fig.~\ref{fig:curvat} shows the integrated curvature distributions obtained  
for the ``tunneling'' perturbation {\bf B}. The left panel presents the data in 
the transition between ergodic and MBL regime. 
The curvature distribution (and its integral{, $F(K)$,} depicted in the figure) retains 
the algebraic tail for large curvatures, $|k|^{-\alpha}$ with, however $\alpha$ 
smoothly changing with increase of $W$. In this regime the data for 
``interaction'' perturbation {\bf A} exactly match those for perturbation {\bf 
B} despite the fact that velocity distributions differ.

The situation changes in deeply localized regime as shown in the right panel 
again for perturbation {\bf B} only. The obtained distributions seem to 
correspond to two types of levels. Levels in the first group are characterized 
by small curvatures -- large curvatures are exponentially avoided. On the other 
hand there remains a fraction of levels with large curvatures and algebraic 
tail of the integrated distribution. {T}he 
corresponding power 
(see dashed lines in the figure) changes smoothly between $0.66\pm0.03$ and $1$ at very large $k$. 
Note, that with increasing $W$ the 
first group grows while the second group of levels shrink. For $W=64$ the 
algebraic behavior corresponds to at most a ``permille'' of all curvatures 
collected.

\begin{figure}[ht]
\includegraphics[width=1.1\linewidth]{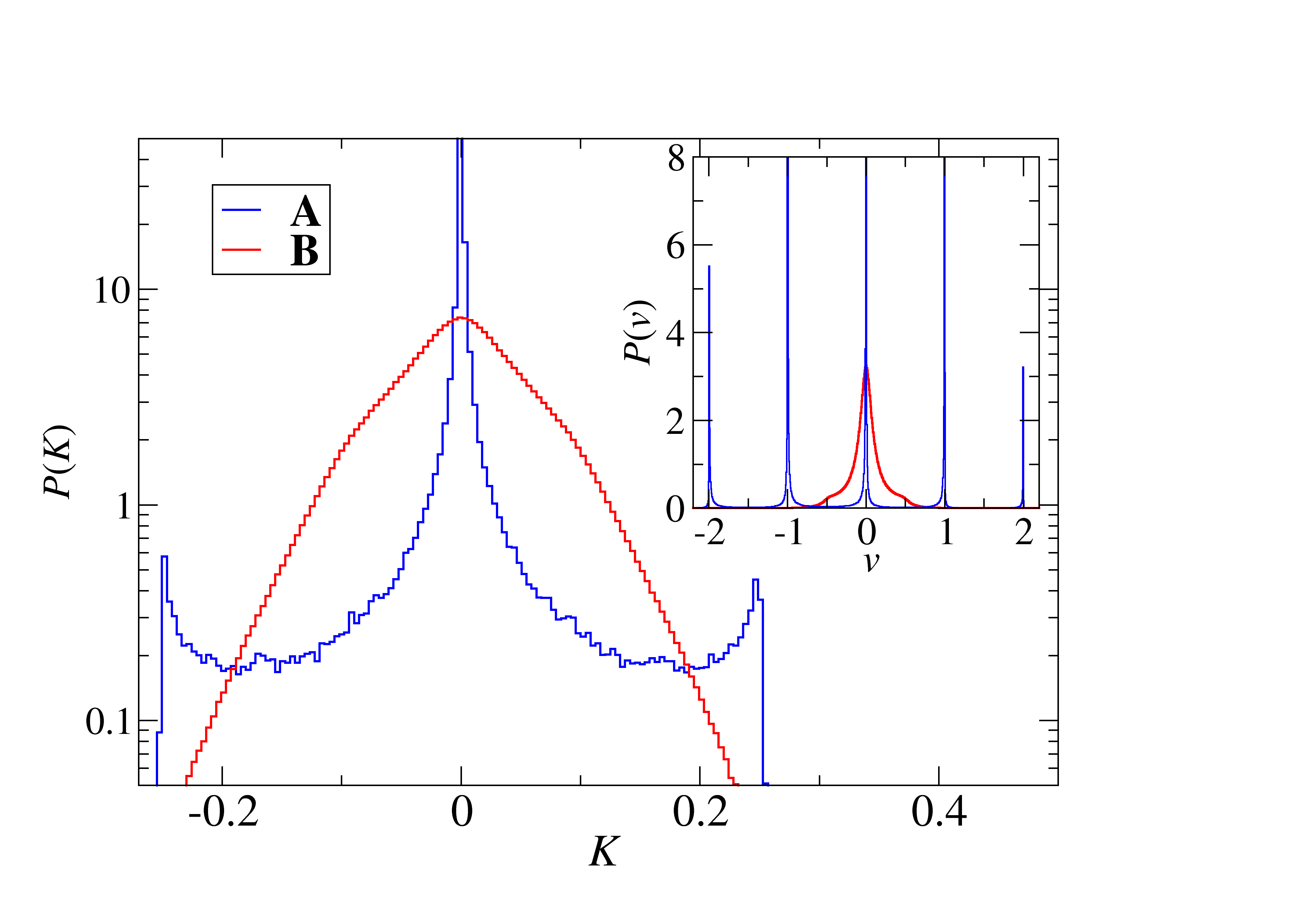} 
	\caption{Unscaled curvature distributions in a deeply localized regime, 
$W=64$,  
for ``interaction'' perturbation {\bf A } -- blue line as compared to 
``tunneling'' perturbation {\bf B} (red line). The unexpected shape of the 
former with the excess peak at $K=0$ -- corresponding to levels with a fixed 
constant slope correlates with the velocity distribution shown in the inset.}
	%	\vspace{-3ex}
	\label{fig:curloc}
\end{figure}

On the other hand the distribution of rescaled curvatures in the localized 
regime for {\bf A} perturbation shows strange irregularities.
Inspection of single realizations reveals that few distinct values of curvatures 
appear, most notably vanishing curvatures are abundant.
This suggests that energy levels form straight lines as a function of the parameter. Moreover, we have seen 
already in Fig.~\ref{fig:curv} that velocities for the type {\bf A} 
perturbations are peaked at distinct values. In effect, the  statistics of 
curvatures for {\bf A} perturbation does not bring other interesting 
information except the fact that levels are organized in groups of very similar 
velocities -- compare the inset in Fig~\ref{fig:curloc}. Levels within the group 
have predominantly very small curvatures as exemplified in the main panel of 
Fig~\ref{fig:curloc}. Only when levels corresponding to different 
velocity groups cross, the resulting narrow avoided crossings lead to the appearance 
of large curvatures as visible in Fig~\ref{fig:curloc} as sidebands in the blue 
curve. This peculiar behavior is due to the fact that the ``direction'' of {\bf 
A} perturbation is along the conserved LIOMs structure. For other more generic 
perturbation,  for instance of {\bf B} type we observe smooth distributions 
of velocities and curvatures (red lines in Fig~\ref{fig:curloc}).

\subsection{Distribution of curvature ratio}
\label{subsec:CR}
We have seen above that the analysis of curvature distributions requires 
scaling of curvatures. Following \cite{Filippone16} we have consistently used 
rescaling of each realization by its width (defined on the basis of cumulative 
curvature distribution). While this procedure is well defined, it is by no means 
unique. Rescaling, e.g. by the typical width obtained averaging over all 
disorder realizations yields different results. Therefore, it is desirable to 
define a measure independent of the rescaling. The situation is, somewhat 
similar to that for level spacings. There, the gap ratio {which avoids unfolding}
is {commonly} used (see 
Introduction). Here we introduce, therefore, the curvature 
ratio, whose distribution, as we shall see, provides additional information 
about the system studied. 
 
The curvature ratio is defined as a ratio of two curvatures for consecutive 
levels, $D_n=K_n/K_{n+1}$. The intuition suggests that $D_n$ may behave 
interestingly in the isolated avoided crossing region where one expects 
$D_n\approx -1$. For ergodic systems one expects relatively large avoided 
crossings \cite{Zakrzewski91,Zakrzewski93c}, levels are affected by many
neighbors and in effect there is smaller direct correlations between consecutive levels.

\begin{figure*}[ht]
\includegraphics[width=0.53\linewidth]{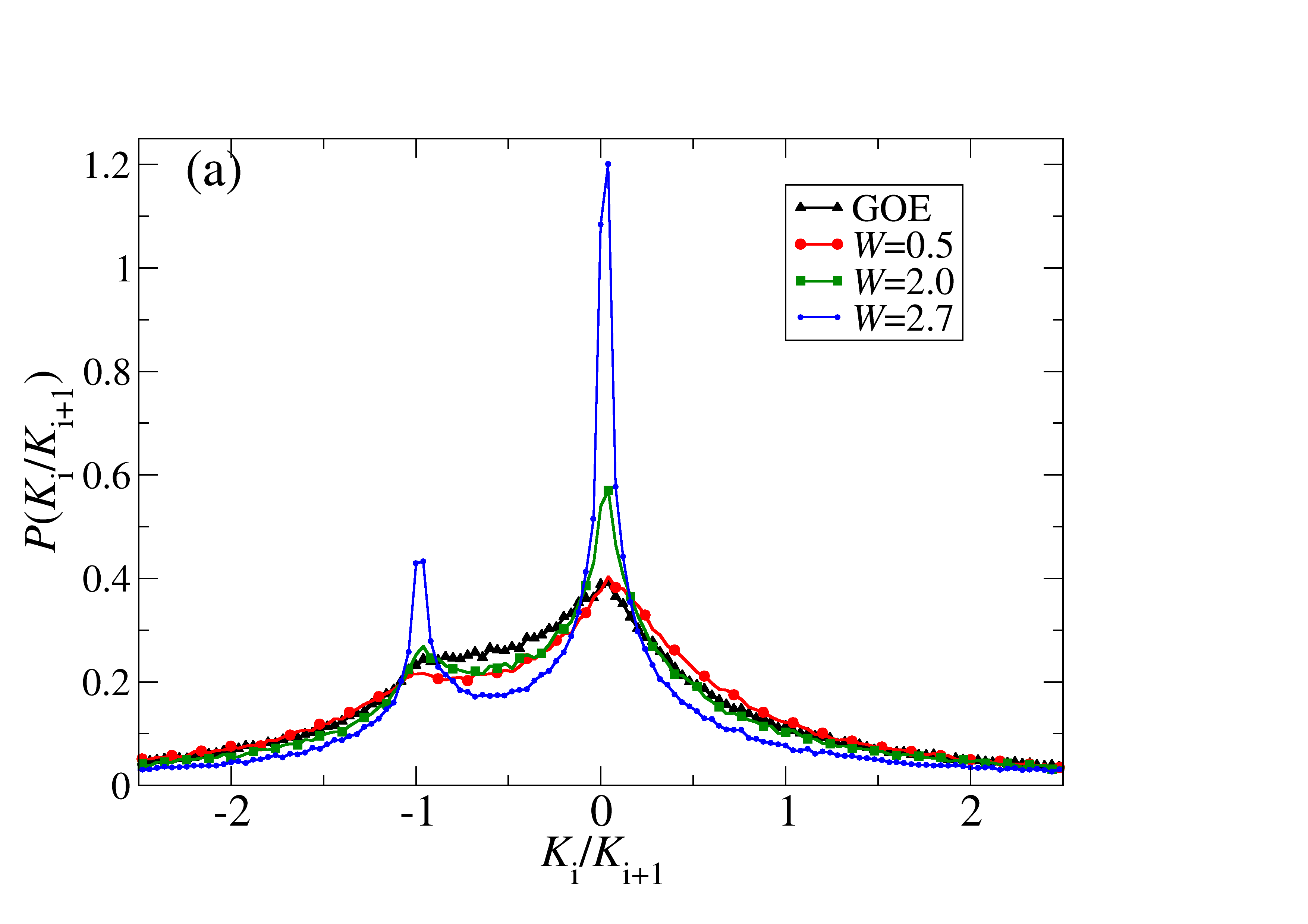}\includegraphics[width=0.53\linewidth]{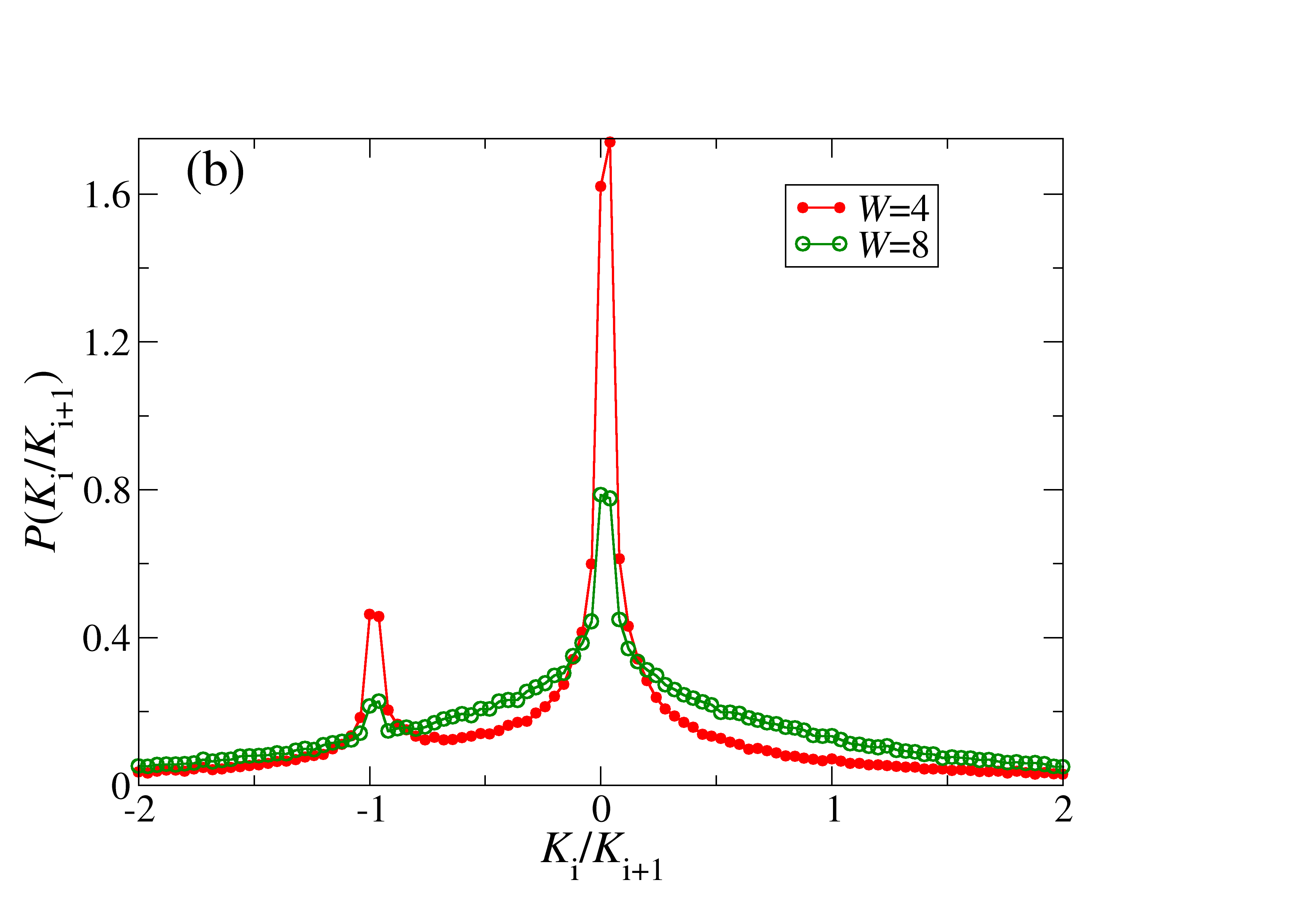} 
	\caption{The distribution of curvatures ratio for delocalized (a) and 
localized phase (b) for the ``tunneling'' perturbation {\bf B}. 	\label{fig:ratio}}
\end{figure*}

The distribution of curvature ratio for the values of disorder that corresponds 
to delocalized phase (Fig.~\ref{fig:ratio}(a)) and to localized system 
(Fig.~\ref{fig:ratio}(b)) is shown in Fig.~\ref{fig:ratio}. 
For small disorder values the distribution of curvature ratio is similar to the 
GOE one, however with increase of the disorder strength, $W$, one can observe 
the appearance of pronounced peaks at the value of $-1$ and $0$. The former 
suggest increasing abundance of curvatures with equal absolute values but 
opposite sign -- that is  clearly related to the importance of avoided crossing 
of energy levels. 
This can be seen from results in Fig.~\ref{fig:ratio}(a) where black line is 
for GOE, for which the peak at $D_n=-1$ is barely visible.
It becomes more important for the transition region $W$ values. Interestingly, 
deep at the localized phase the peak diminishes indicating 
less abundant avoided crossings.

It is worth to note that
even more significant seems the peak at $D_n\approx0$. This 
corresponds to events where a level with small curvature (constant slope) 
becomes close to other curved level. Levels of constant slope are a natural 
candidates for the indicator of LIOMs. Why this peak becomes also less 
pronounced for large $W$? This may be due to the fact that close levels with 
similar slope and small curvatures dominate the spectrum, then the ratio of 
curvatures becomes less sensitive to extreme values.

\subsection{Fidelity susceptibility distributions}

Last, but not least, let us inspect the fidelity susceptibility distributions 
in the transition to MBL and in the deeply localized regime. We consider all 
three different perturbation schemes.

Let us first discuss the transition to MBL, this time in the ``tunneling'' 
parametric level dynamics, i.e. model {\bf B}.
\begin{figure*}[ht]														
\includegraphics[width=0.53\linewidth]{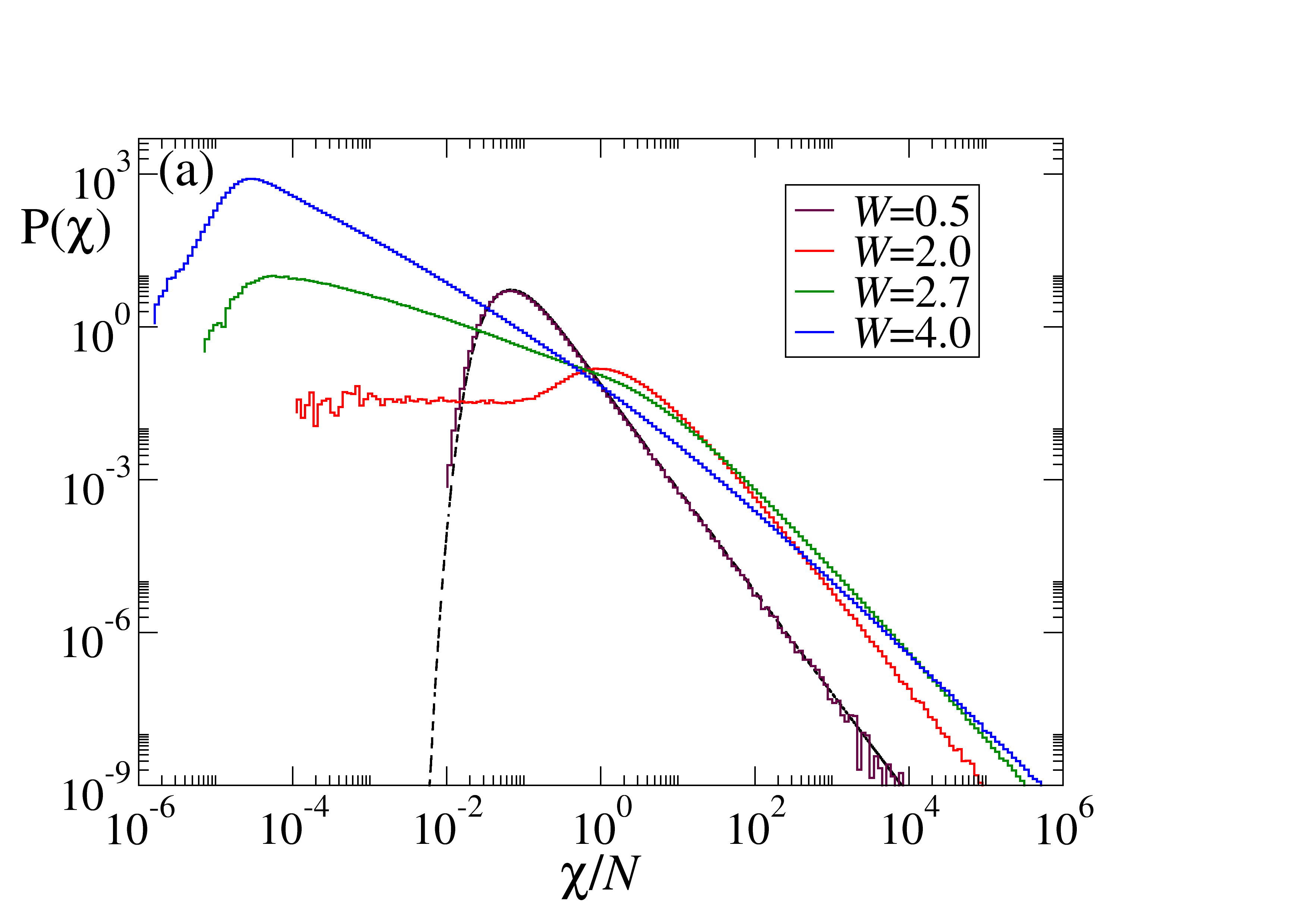}\includegraphics[width=0.53\linewidth]{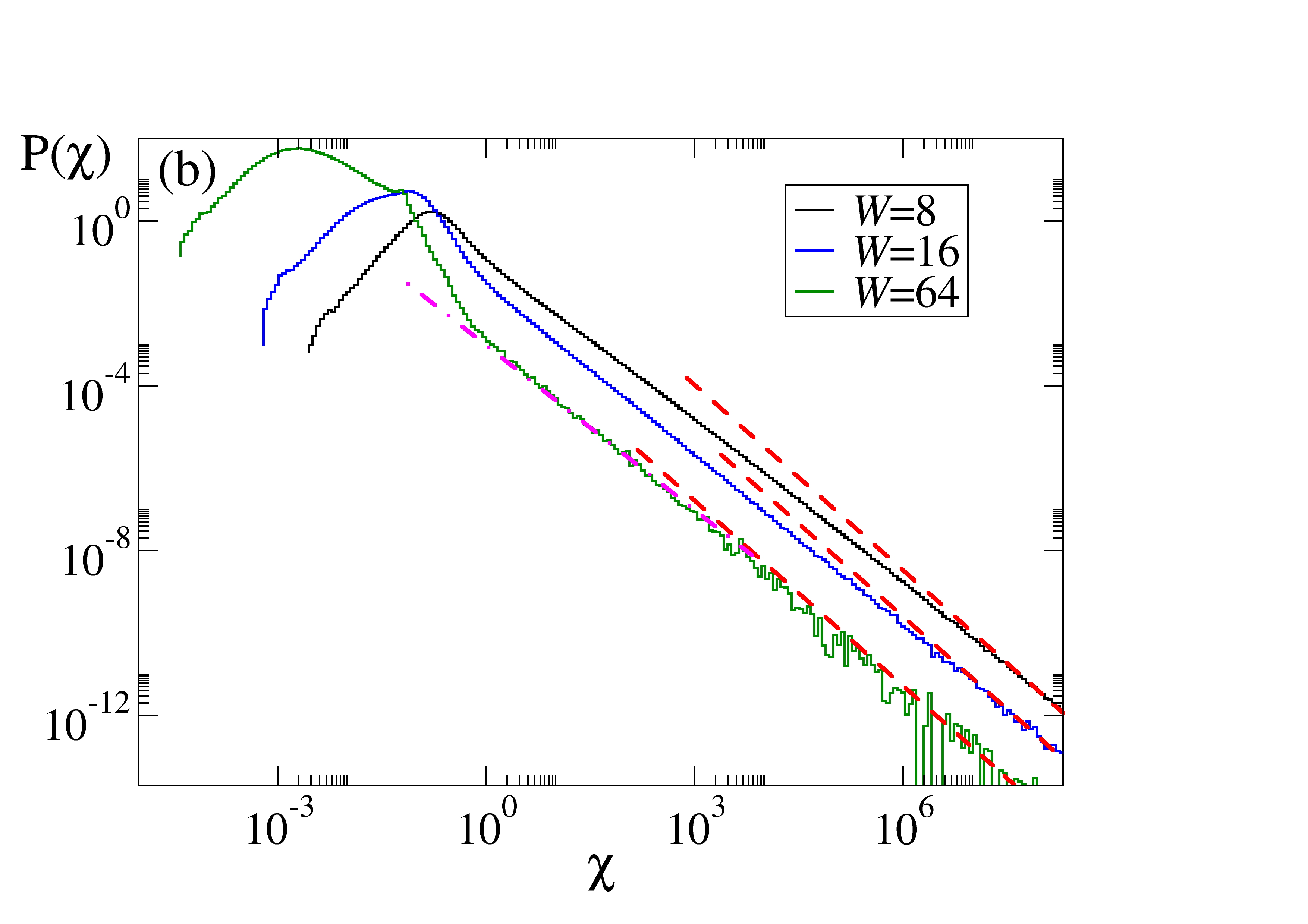} 
	\caption{Fidelity susceptibility distribution for ``tunneling'' 
perturbation {\bf B} in the transition to localized (a) and deeply in  the 
localized regime (b). The primary feature is the appearance of very small 
fidelity susceptibilities, characteristic for localized levels. The large 
fidelity {tail is} $\chi^{-2}$ in the delocalized 
{regime}. {I}n 
the deeply localized regime, a power law-decay with exponent 
varying between $-4/3$ and  $-3/2$ for extremely large $\chi$, is visible
 being there only weakly dependent on the disorder 
strength $W$. For deeply localized regime observe a kink at  $\chi\approx 0.2$.\label{fig:fidxy}}
\end{figure*}
Fig.~\ref{fig:fidxy} shows the change in the fidelity susceptibility 
distribution with increasing $W$ i.e. entering the localization regime.
While for pure ergodic behavior (black line for $W=0.5$ with dashed line 
reproducing analytic expression (\ref{pchi})) small susceptibilities are 
strongly avoided, this is not the case for large disorder $W$. Upon entering 
the crossover regime (remember we consider $L=16$ case) very small 
susceptibilities become abundant, at the same time a slope for large 
susceptibilities changes. In the localized regime, as depicted in 
Fig.~\ref{fig:fidxy}(b) the large fidelity susceptibility {tail can be 
locally described as a power law with exponent $\alpha$ changing smoothly from 
$-1.33\pm0.04$ to $-3/2$ for tiny fraction of $\chi \gtrsim 10^6$.}
{This tail behavior is relevant for a very }
small fraction of levels, the most of them {are characterized by} low susceptibilities. Note that 
these data correlate quite well with those for curvatures in 
Fig.~\ref{fig:curvat}(b).

\begin{figure*}[ht]
\hfill
		\includegraphics[width=0.53\linewidth]{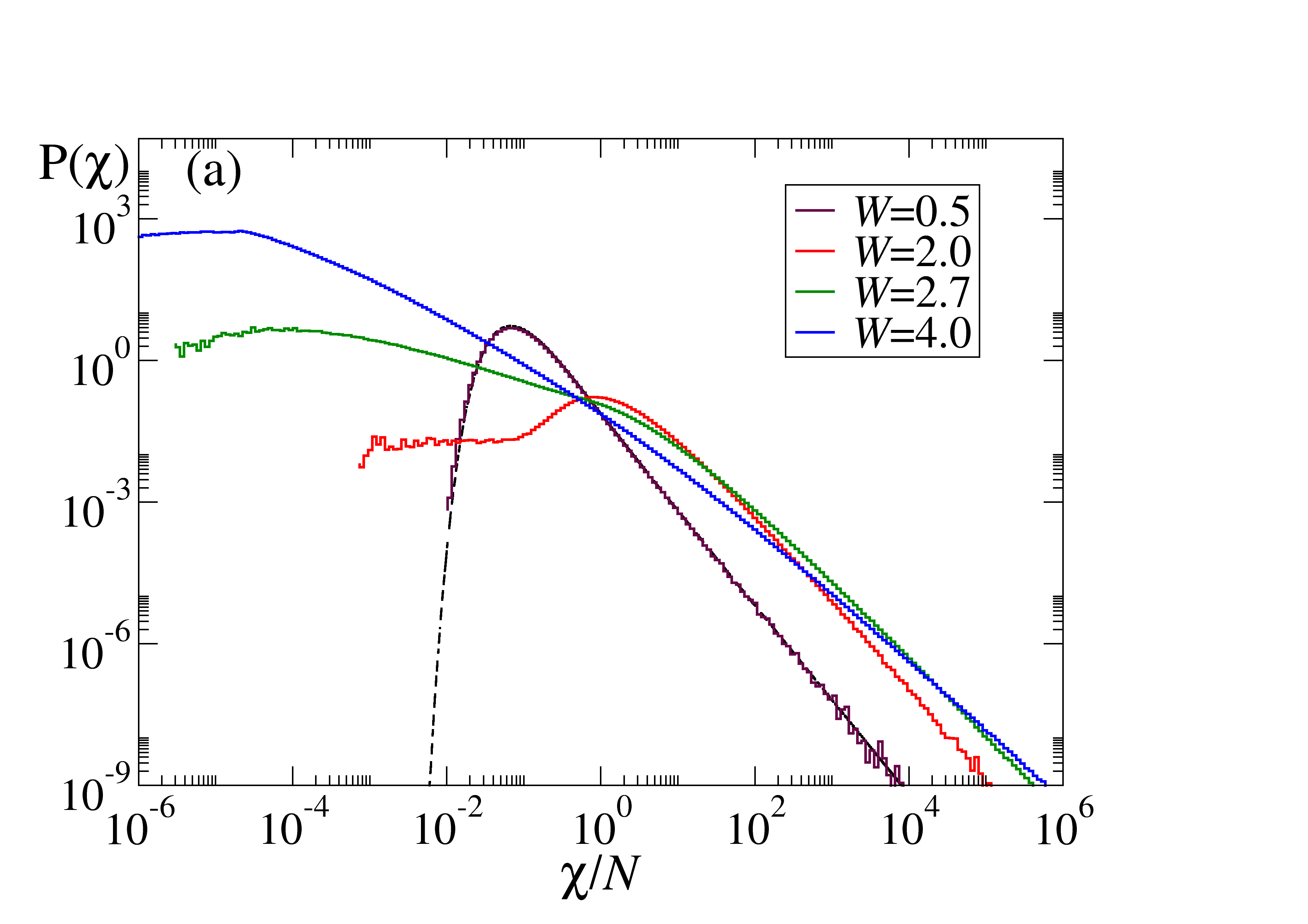}\includegraphics[width=0.53\linewidth]{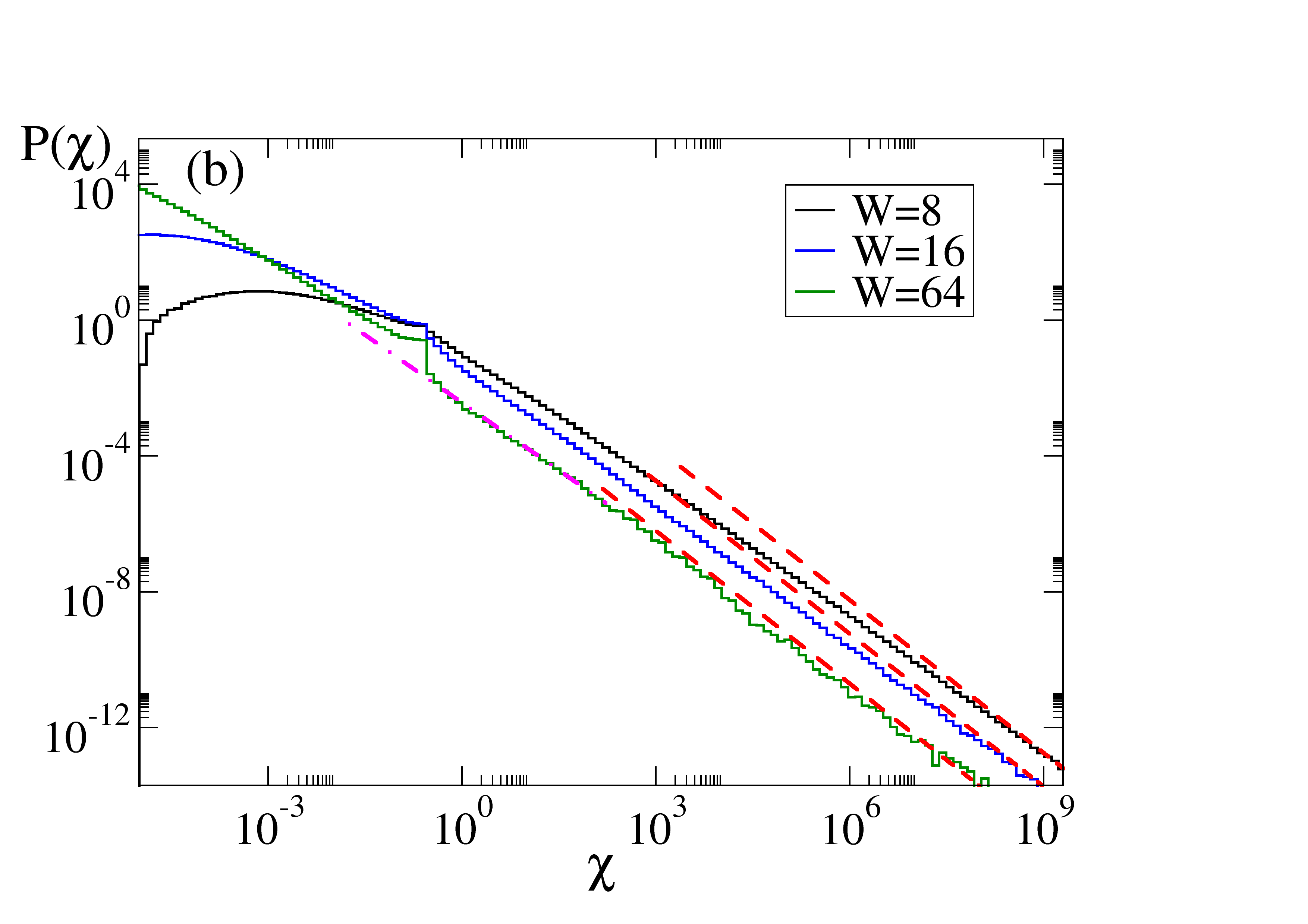} 
	\caption{Same as Fig.~\ref{fig:fidxy} but for the ``interaction'' 
perturbation {\bf A}. Panel (a) shows data for $L=16$ in the transition regime, 
panel (b) in the deeply localized region. As before, observe a kink at 
$\chi\approx 0.2$ in panel (b).}
	%	\vspace{-3ex}
	\label{fig:fidz}
\end{figure*}
A quite similar picture is obtained for the perturbation {\bf A}. In the 
transition regime distributions for both {\bf A} and {\bf B} perturbations seem 
very similar. This picture changes in the localized regime as observed for 
$W=4$ in panel (a) of Fig.~\ref{fig:fidxy} and Fig.~\ref{fig:fidz} as well as in 
the right hand panels in both figures. While for perturbation {\bf B} 
``perpendicular'' to LIOMs the smallest fidelity susceptibilities are avoided, 
this is not the case for perturbation {\bf A} ``parallel'' to LIOMs -- here 
small fidelities are most abundant. On the other hand largest fidelity 
susceptibilities again show (compare Fig.~\ref{fig:fidz}) the power law decay 
with the slope {changing from} about $-4/3$ 
{to $-3/2$} (here fluctuations related to the particular value 
of $W$ are slightly bigger than for the ``tunneling'' perturbation {\bf B}). 

\begin{figure}[ht]
		\hfill{\includegraphics[width=1.1\linewidth]{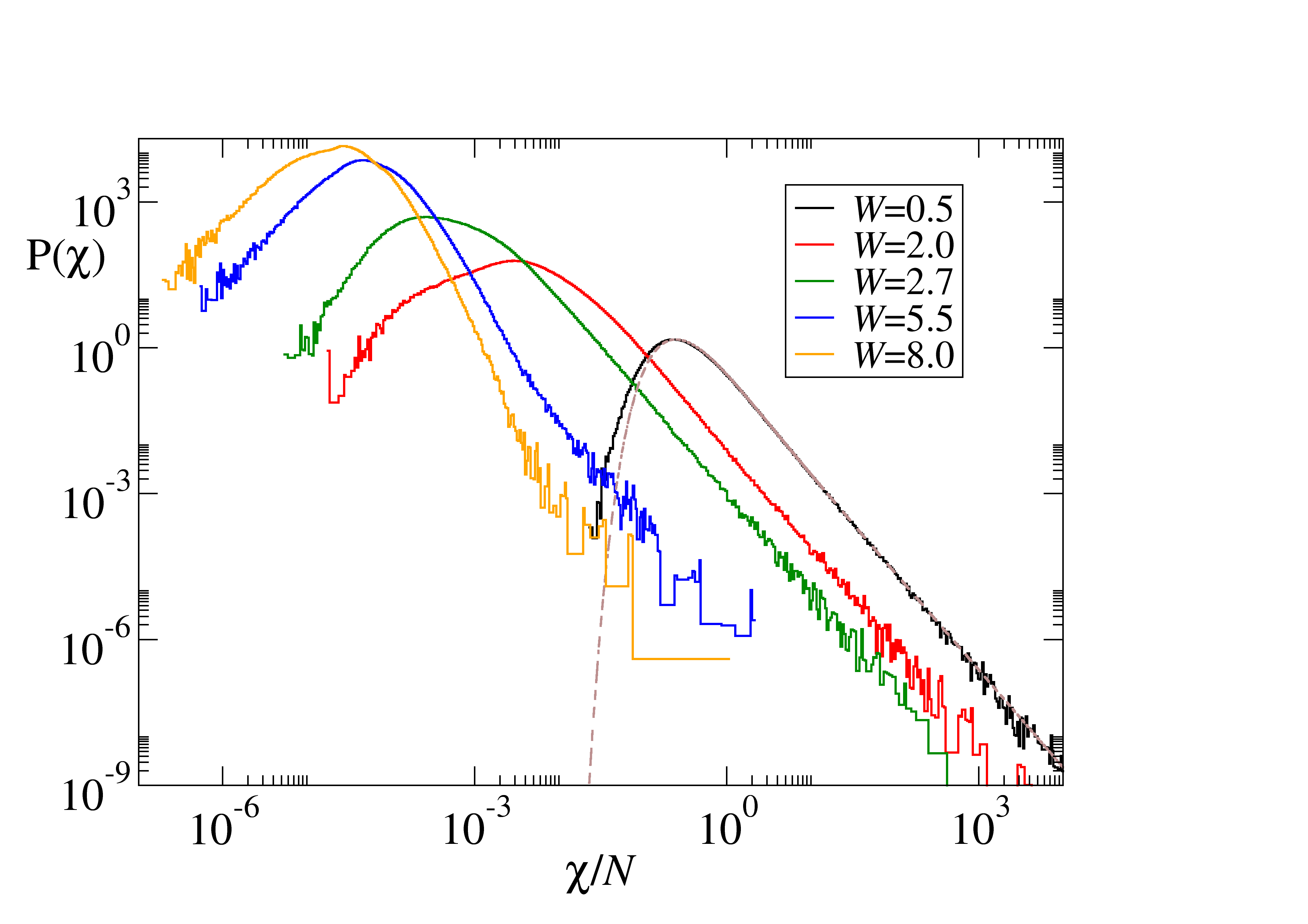} } 
	\caption{Fidelity susceptibility distribution for infinitesimal 
Aharonov-Bohm flux case. In the delocalized regime the distribution is 
different from the standard GOE case for small values of susceptibility. For 
bigger disorder strength, in the MBL regime, a definite lack of large 
susceptibilities (as compared to other perturbations) is observed. }
	%	\vspace{-3ex}
	\label{fidvO}
\end{figure}
Finally let us consider the case of Aharonov-Bohm flux perturbation {\bf C} -- 
compare Fig.~\ref{fidvO}. The main difference with other perturbations that 
preserve time reversal invariance lies for small fidelity susceptibilities. Those are strongly 
avoided, although the maximum of the distribution moves to smaller $\chi$ the 
stronger the disorder $W$ is. More importantly, in the MBL regime, the 
distribution of fidelity susceptibility does not seem to follow a power law for 
large susceptibilities. The decay is much faster. Only for the largest $\chi$'s 
(for a given $W$) one may observe the remnants of the algebraic tail. 
Interestingly its slope is again, to a good accuracy $\chi^{-2}$, as in the 
delocalized regime. Thus it seems
that this tail is due to some rare delocalized events (Griffiths regions) due, 
most presumably, to the size effects.

\section{Discussion}
\label{disc}

The numerical results presented above call for the summarizing discussion. 
Firstly, both in the delocalized and in MBL regimes the level dynamics is not 
universal in the system studied, the XXZ Hamiltonian, a paradigmatic system for 
MBL studies.  It is particularly surprising that this nonuniversality appears 
also in the ergodic regime where predictions of RMT should hold. We believe 
that this in not a finite size effect. While we have limited our study to system 
of size $L=16$ and smaller, we have checked that, e.g. velocities for $L=18$ 
behave similarly to $L=16$. Let us stress again that the observed nonuniversality in the delocalized
regime is to some extent in contradiction with purely ergodic behaviour in this parameter
regime reported on the basis of participation ratio studies \cite{Mace18}. 
Additional studies, in particular of the distribution of participation ratios not just their mean value,
may be needed to clarify fully this issue.

We have confirmed that, in the delocalized regime, curvature distribution 
faithfully obeys (\ref{probRMT}) provided the curvatures are scaled by the 
width extracted from the integrated distribution in a way proposed by 
\cite{Filippone16}. Then the same distribution holds both for perturbations 
within time reversal invariant  system class  as well as for an infinitesimal Aharonov-Bohm flux breaking 
that symmetry.  In a similar way, the fidelity susceptibility distribution in 
the ergodic regime is faithfully reproduced by the recently found analytic 
prediction for GOE \cite{Sierant19} with small deviations observed for the 
Aharonov-Bohm flux case. 

The lack of universality in level dynamics can be intuitively understood in the following
way. The expression \eqref{curv} for level curvature $K_n$ contains the 
sum of terms, each of them proportional to the appropriate off-diagonal
matrix element of the perturbation matrix $|\langle \psi_k|V|\psi_n\rangle|^2$. 
To see the universality one rescales the $K_n$ by variance of the velocity distribution. 
Velocities, in turn, are diagonal matrix elements $\langle \psi_n|V|\psi_n\rangle$.
Therefore, if, in a quantum chaotic system the ratio of off-diagonal to diagonal matrix elements
of perturbation is the same as for GOE, then one obtains the universality of level dynamics.
If, on the other hand, the ratio is different than for GOE, the universality is broken. This 
is the case for the studied XXZ spin chain, the paradigmatic model of MBL. 

At the other extreme, deeply in the localized regime we find that the shape of 
distributions for velocities, curvatures and fidelities strongly depend on the 
perturbation assumed. The ``interaction'' perturbation {\bf A} is quite 
peculiar as it is almost diagonal in the basis of LIOMs leading to unusual 
velocity distributions -- the level slopes become approximately quantized. 
Due to the simplicity of XXZ model and its structure of LIOMs being almost diagonal
in eigenbasis of $S^z_i$ operators, 
such a clear difference between perturbations {\bf A} and {\bf B} 
appears. This suggests that velocity distributions
studied for different perturbations may turn out to be
quite useful for identifying the structure and properties of LIOMs for more complicated
geometries and systems with lack of  strong localization on physical sites.
  
Surprisingly, curvature distributions seem less sensitive to direction of perturbation at least across the transition to
MBL regime. For moderate $W$, even on localized side, the curvature distribution for 
perturbation {\bf A} is very similar to that for a ``generic'' tunneling 
perturbation {\bf B}. The latter yields Gaussian distribution of velocities and 
curvature distributions that smoothly evolve into the MBL regime. 
The power law tail of curvature (fidelity susceptibility)
distribution agrees  with the 
predictions of \cite{Monthus17} which are $k^{-2}$ ($\chi^{-3/2}$)
only for extremely large curvatures (fidelities). 
This shows that the assumption inherent in
\cite{Monthus17} that the matrix elements of the perturbation operator  in the 
numerator of (\ref{curv}) or (\ref{fs}), $|\langle \psi_k | V | 
\psi_n\rangle|^2$ are independent from powers of level spacing $E_n-E_k$ in the 
denominator is fulfilled only in very rare, extremal cases, being at the same
time a plausible assumption for Gaussian random matrices.

Similar situation occurs already for noninteracting particles in the Anderson 
regime. The contributions  to large curvature (fidelity) tail correspond 
necessarily to  almost degenerate  levels $E_n$ and $E_k$ in the sums in 
(\ref{curv}) or (\ref{fs}). Such levels must be 
decorrelated (remember Poisson level spacings) with wavefunctions localized 
(exponentially) in different regions in space. Thus the matrix element of the 
local operator in the numerator should also generically decay exponentially. 
This leads to e.g. log-normal prediction for curvature distribution in deeply 
localized regime \cite{Titov97}. 
Apparently the situation is more subtle in the MBL case, the power law tails 
are preserved. 

The observed features of level dynamics vary continuously in ergodic-MBL crossover regime,
unexpectedly, some changes are visible even deep in the MBL phase where the level statistics 
is purely Poissonian.
While close to the transition region the most prominent feature of the
curvature distribution, the large curvature tail changes smoothly with the 
disorder amplitude,
for large disorder a clear difference occurs.  The majority of levels strongly avoids
large curvatures, the algebraic tail is visible for a tiny fraction of levels (c.f. Fig.~\ref{fig:curvat}). 
Again, only in this deeply localized regime different perturbations affect strongly the curvature distribution
due to the geometry of LIOMs.

\section{Conclusions}
\label{conc}

We have analyzed, mostly numerically, the level dynamics for a system 
undergoing ergodic to MBL phase transition. Most surprisingly we have found that 
the level dynamics does not obey commonly believed universality as expressed in 
the seminal paper of Simons and Altschuler \cite{Simons93}. At the same time we 
have found that a simple rescaling allows to fit  RMT-based well known 
expression for the curvatures \cite{Zakrzewski93} as well as the recently found 
large size limit of the fidelity susceptibility distribution \cite{Sierant19}.
{Moreover, our results show deviations between GOE level dynamics
and level dynamics of an ergodic system, indicating to what extent
an ergodic system can be modeled by a GOE matrix.}

Upon transition to MBL and in the MBL regime we have found that level dynamics 
is dependent on the character of the perturbation. 
For ``interaction'' perturbation almost diagonal in LIOMs basis the velocities 
become effectively quantized. More generic ``tunneling'' perturbation yields 
Gaussian distribution of slopes of energy levels with corresponding fidelity 
susceptibilities as well as curvatures decaying algebraically in the large 
value limit.  This is not the case for the infinitesimal Aharonov-Bohm flux 
perturbation which, while breaking time-reversal invariance of the system, 
suppresses large curvatures and fidelities.

This work paves the way to more complete and detailed analysis of level 
dynamics. On one hand, one may compare the effects due to uniform random 
disorder considered in this work with those obtained with quasi-periodic 
disorder as realized in experiments \cite{Schreiber15}. On the other hand one 
may consider different perturbations of the system. While we considered 
``global'' perturbations in the form of sum over sites of local observables -- 
one may consider purely local cases \cite{Serbyn15}. Last but not least -- the present 
study was necessarily limited to a single system - the XXZ model. A similar analysis for 
the Ising models or schemes beyond nearest-neighbor tunnelings/interactions may
verify, for example, the extend to which the universality of level dynamics is
destroyed in many body systems.
The possible differences 
may help us to understand the intricacies of many body localization phenomenon.

\section{Acknowledgement} 
We are grateful to Dominique Delande and Wojciech de Roeck for helpful 
discussions throughout the course of this work.  
%We acknowledge also useful conversations with Fabian Alet and Antonello 
%Scardicchio. 
This research was performed within 
projects   No.2015/19/B/ST2/01028 and 2018/28/T/ST2/00401 (P.S -- doctoral scholarship)
financed by  National Science Centre (Poland). 
QuantERA QTFLAG
programme No. 2017/25/Z/ST2/03029 (J.Z.) implemented within the European 
Union's Horizon 2020 Programme is also acknowledged as well as the important 
support by PL-Grid Infrastructure.

%\section{References}
%\bibliographystyle{unsrt}
% \bibliography{ref2018v1}

\end{document}